\documentclass[superscriptaddress,
reprint,longbibliography,
 amsmath,amssymb,
aps,
prb,
]{revtex4-1}

\usepackage{simpler-wick}
\usepackage{graphicx}
\usepackage{dcolumn}
\usepackage{bm}
\usepackage{subfig}

\usepackage[colorlinks=true,linkcolor=blue,citecolor=blue,urlcolor=blue]{hyperref}
\usepackage{braket}

\DeclareMathOperator{\tr}{tr}

\DeclareMathOperator{\HF}{HF}
\DeclareMathOperator{\QP}{QP}

\usepackage{bbm}

\newcommand{\h}{\mathrm{h}}
\newcommand{\p}{\mathrm{p}}
\newcommand{\CC}{\mathrm{CC}}

\newcommand{\sat}{\mathrm{sat}}
\newcommand{\F}{\mathrm{F}}
\newcommand{\B}{\mathrm{B}}

\usepackage{feynmf}

\begin{document}


\title{Uncovering relationships between the electronic self-energy and coupled-cluster doubles theory}

\author{Christopher J. N. Coveney}
\email{christopher.coveney@physics.ox.ac.uk}
\affiliation{ Department of Physics, University of Oxford, Oxford OX1 3PJ, United Kingdom}

\date{\today}

\begin{abstract}
\begin{center}
    \textbf{Abstract}
\end{center}
    We derive the coupled-cluster doubles (CCD) amplitude equations by introduction of the particle-hole-time decoupled electronic self-energy. The resulting analysis leads to an expression for the ground state correlation energy that is exactly of the form obtained in coupled-cluster doubles theory. We demonstrate the relationship to the ionization potential/electron affinity equation-of-motion coupled-cluster doubles (IP/EA-EOM-CCD) eigenvalue problem by coupling the reverse-time self-energy contributions while maintaining particle-hole separability. The formal relationships established are demonstrated by exact solution of the Hubbard dimer. 
\end{abstract}


\maketitle

\begin{fmffile}{diagram}

\section{Introduction}

Green's function methods simultaneously encode excited and ground state many-body correlation. This is reflected by the fact that exact ground state properties as well as the single-particle charged excitation spectrum can be obtained from the same single-particle Green's function.~\cite{Quantum,stefanucci2013nonequilibrium,mahan2000many} Recently there has been significant interest in identifying and leveraging connections between coupled-cluster and Green's function theory.~\cite{quintero2022connections,monino2023connections,berkelbach2018communication,tolle2023exact,bintrim2021full,bintrim2022full,lange2018relation,opoku2021new,opoku2023new,opoku2023new_prop,opoku2024new_prop,coveney2023coupled,coveney2024cc_se} In this work, we uncover the connection between the electronic self-energy and the coupled-cluster doubles amplitude equations by showing how, under certain approximations, the `upfolded' quasiparticle equation can be recast to give the CCD amplitude equations. In the following, indices $i,j,k,\cdots$ denote occupied (valence band) spin-orbitals, $a,b,c,\cdots$ virtuals (conduction band) and $p,q,r,\cdots$ general spin-orbitals.  We employ a real canonical spin-orbital basis throughout this work.

\section{Green's function theory and the Algebraic Diagrammatic Construction method}

The central quantity in the perturbation theory of the single-particle Green's function is the self-energy. Through the Dyson equation, an approximation for the self-energy defines the corresponding approximation for the single-particle Green's function. The electronic self-energy has a spectral representation that is a consequence of the analytic structure of the single-particle Green's function and is given by~\cite{schirmer1983new,caruso2013self,raimondi2018algebraic,schirmer2018many,coveney2023coupled,coveney2024cc_se}
\begin{gather}
    \begin{split}~\label{eq:spec_se}
        \Sigma_{pq}(\omega) &=\Sigma^{\infty}_{pq}+\Sigma^{\F}_{pq}(\omega)+\Sigma^{\B}_{pq}(\omega)\\
        &= \Sigma^{\infty}_{pq} + \sum_{JJ'} U^\dag_{pJ}\left[(\omega+i\eta)\mathbbm{1}-(\mathbf{K}^{>}+\mathbf{C}^{>})\right]^{-1}_{JJ'}U_{J'q}\\
        &+ \sum_{AA'} V_{pA}\left[(\omega-i\eta)\mathbbm{1}-(\mathbf{K}^{<}+\mathbf{C}^{<})\right]^{-1}_{AA'}V^\dag_{A'q} \ .
    \end{split}
\end{gather}
$\Sigma^{\infty}_{pq}$ is the static, frequency-independent contribution to the electronic self-energy and $\Sigma^{\F/\B}_{pq}(\omega)$ are the dynamical forward-/backward-time self-energy contributions, respectively. From the spectral representation of the self-energy, it can be shown that the \emph{eigenvalue}-self-consistent frequency-dependent quasiparticle equation is equivalent to diagonalization of the `upfolded' frequency-independent Dyson supermatrix~\cite{backhouse2020wave,bintrim2021full,scott2023moment,raimondi2018algebraic,schirmer2018many,coveney2023coupled,coveney2024cc_se}
\begin{gather}
    \begin{split}~\label{eq:el_dys}
        \mathbf{D} = \left(\begin{array}{ccc}
            f_{pq}+\Sigma^{\infty}_{pq} & U^\dag_{pJ'} & V_{pA'} \\
            U_{Jq} & (\mathbf{K}^{>}_{JJ'}+\mathbf{C}^{>}_{JJ'}) & \mathbf{0} \\
             V^\dag_{Aq} & \mathbf{0} & (\mathbf{K}^{<}_{AA'}+\mathbf{C}^{<}_{AA'}) 
        \end{array}\right) .
    \end{split}
\end{gather}
The composite indices $JJ'/AA'$ label forward-time/backward-time Intermediate State Configurations (ISCs) and outline the character of the different multi-particle-hole configurations that connect the initial and final Green's function spin-orbital indices as a result of interaction and particle propagation in the system. ISCs are excited state configurations containing $(N \pm 1)$-electrons that 
can be related to specific electronic configurations resulting from excitation with respect to a general reference state.~\cite{schirmer1983new,mertins1996algebraicI,mertins1996algebraicII,raimondi2018algebraic,coveney2023coupled,coveney2024cc_se} The matrices ($\mathbf{K}^{>}_{JJ'}+\mathbf{C}^{>}_{JJ'}$) and ($\mathbf{K}^{<}_{AA'}+\mathbf{C}^{<}_{AA'}$) represent the interactions between the different ISCs. The quantities $U_{qJ}$ and $V_{pA}$ represent the coupling matrices that link initial and final single-particle Green's function indices to the ISCs. The upper left block of $\mathbf{D}$ is defined over the complete set of occupied and virtual spin-orbitals, where $f_{pq}=\epsilon_{p}\delta_{pq}$ is the Fock operator. The coupling and interaction matrices contain coupling to all possible ISCs. The zero entries of the Dyson supermatrix are present as the forward-and backward-time self-energy contributions are coupled only through the initial and final single-particle Green's function indices. Diagonalization of the Dyson supermatrix yields the complete set of ionization potentials and electron affinities:
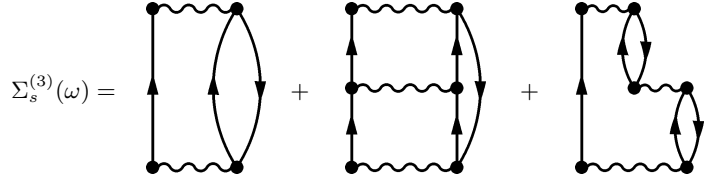
\begin{figure*}[ht]
    \centering
    \begin{gather*}
\begin{split}~\label{eq:cc_II_pt}
\Sigma_{s}^{(3)}(\omega) &= \hspace{2.5mm}
\begin{gathered}
    \begin{fmfgraph*}(40,60)
    \fmfcurved
    \fmfset{arrow_len}{3mm}
    \fmfleft{i1,i2}
    \fmflabel{}{i1}
    \fmflabel{}{i2}
    \fmfright{o1,o2}
    \fmflabel{}{o1}
    \fmflabel{}{o2}
    \fmf{fermion}{i1,i2}
    \fmf{wiggly}{o1,i1}
    \fmf{fermion,left=0.3,tension=0}{o1,o2}
    \fmf{wiggly}{o2,i2}
    \fmf{fermion,left=0.3,tension=0}{o2,o1}
    \fmfdot{o1,o2,i1,i2}
\end{fmfgraph*}
\end{gathered} \hspace{5mm}+\hspace{5mm} 
    \begin{gathered}
    \begin{fmfgraph*}(40,60)
    \fmfcurved
    \fmfset{arrow_len}{3mm}
    \fmfleft{i1,i2,i3}
    \fmflabel{}{i1}
    \fmflabel{}{i2}
    \fmfright{o1,o2,o3}
    \fmflabel{}{o1}
    \fmflabel{}{o2}
    \fmf{wiggly}{i2,o2}
    \fmf{wiggly}{i3,o3}
    \fmf{wiggly}{i1,o1}
    \fmf{fermion}{i1,i2}
    \fmf{fermion}{i2,i3}
    \fmf{fermion}{o1,o2}
    \fmf{fermion}{o2,o3}
    \fmf{fermion,left=0.3}{o3,o1}
    \fmfforce{(0.0w,0.0h)}{i1}
    \fmfforce{(1.0w,0.0h)}{o1}
    \fmfforce{(0.0w,1.0h)}{i3}
    \fmfforce{(1.0w,1.0h)}{o3}
    \fmfdot{i1,i2,i3}
    \fmfdot{o1,o2,o3}
\end{fmfgraph*}
\end{gathered}
\hspace{7.5mm}+\hspace{5mm}
\begin{gathered}
\begin{fmfgraph*}(40,60)
    \fmfcurved
    \fmfset{arrow_len}{3mm}
    \fmfleft{i1,i2,i3}
    \fmflabel{}{i1}
    \fmflabel{}{i2}
    \fmfright{o1,o2,o3}
    \fmflabel{}{o1}
    \fmflabel{}{o2}
    \fmf{wiggly}{i1,v1}
    \fmf{wiggly}{v1,o1}
    \fmf{wiggly}{v2,o2}
    \fmf{wiggly}{i3,v3}
    \fmf{fermion}{i1,i3}
    \fmf{fermion,left=0.3}{o1,o2}
    \fmf{fermion,left=0.3}{o2,o1}
    \fmf{fermion,left=0.3}{v2,v3}
    \fmf{fermion,left=0.3}{v3,v2}
    \fmfforce{(0.0w,0.0h)}{i1}
    \fmfforce{(1.0w,0.0h)}{o1}
    \fmfforce{(0.5w,0.5h)}{v2}
    \fmfforce{(0.5w,0.0h)}{v1}
    \fmfforce{(0.5w,1.0h)}{v3}
    \fmfforce{(0.0w,1.0h)}{i3}
    \fmfforce{(1.0w,1.0h)}{o3}
    \fmfdot{v2,v3}
    \fmfdot{i1,i3}
    \fmfdot{o1,o2}
\end{fmfgraph*}
\end{gathered}
\end{split}
\end{gather*}
\caption{The third-order one-particle irreducible skeleton electronic self-energy diagrams. The ADC(3) Dyson supermatrix performs an infinite-order summation of these terms.}
    \label{fig:3rd_order}
\end{figure*}
\begin{gather}
    \begin{split}
        \mathbf{D}\left(\begin{array}{c}
             A^{\nu}_{p}  \\
             W^{+,\nu}_{J} \\
             W^{-,\nu}_{A}
        \end{array}\right) = \left(\begin{array}{c}
             A^{\nu}_{p}  \\
             W^{+,\nu}_{J} \\
             W^{-,\nu}_{A}
        \end{array}\right) \varepsilon_{\nu} \ ,
    \end{split}
\end{gather}
where $\varepsilon_{\nu}$ are the exact poles of the Green's function.
The inverse of the norm of the eigenvectors yields the quasiparticle renormalization factor as~\cite{coveney2024cc_se}
\begin{gather}
    \begin{split}~\label{eq:renorm}
        Z_{\nu}&=\Bigg(\sum_{p}A^{\nu *}_{p} A^{\nu}_p + \sum_{J} W^{+,\nu *}_{J}W^{+,\nu}_{J} +  \sum_{A} W^{-,\nu *}_{A}W^{-,\nu}_{A}\Bigg)^{-1} \\
        &=\Bigg(1-\frac{\partial\Sigma_{\nu\nu}(\omega)}{\partial\omega}\Bigg|_{\varepsilon_{\nu}}\Bigg)^{-1} \ ,
    \end{split}
\end{gather}
where we have implicitly enforced the normalization condition: $\sum_{p}A^{\nu *}_{p} A^{\nu}_p=1$. The sum rule is given by $\sum_{\nu}Z_{\nu}=N_e$, where $N_e$ is the number of electrons.~\cite{von1996self,di2021scrutinizing}   

The third-order Algebraic Diagrammatic Construction method, ADC(3), provides an infinite-order partial summation of a third-order diagrammatic perturbation expansion of the electronic self-energy.~\cite{schirmer1983new,raimondi2018algebraic} It combines diagonalization of the Dyson supermatrix with perturbation expansions of the coupling and interaction matrix elements of the self-energy. For a complete account of the ADC$(n)$ method, the reader is referred to Refs~\citenum{schirmer1983new,schirmer2018many,raimondi2018algebraic}. In this work we simply state the results. The forward-time coupling matrices are defined as 
\begin{subequations}
    \begin{gather}
    \begin{split}~\label{eq:int_forward}
        U^\dag_{p,iab} &= \braket{pi||ab} +\frac{1}{2}\sum_{kl}\braket{pi||kl}(t^{ab}_{kl})_{\text{MP$2$}} \\
        &- \sum_{kc}\braket{pc||bk}(t^{ac}_{ki})_{\text{MP$2$}}+\sum_{kc}\braket{pc||ak}(t^{bc}_{ki})_{\text{MP$2$}} \ ,
    \end{split}
\end{gather}
with the backward-time contributions as
\begin{gather}
    \begin{split}~\label{eq:int_backward}
        V_{p,ija} &= \braket{pa||ij} +\frac{1}{2}\sum_{bc} \braket{pa||bc}(t^{bc}_{ij})_{\text{MP$2$}}\\
        &-\sum_{kb}\braket{pk||jb}(t^{ba}_{ki})_{\text{MP$2$}} +\sum_{kb}\braket{pk||ib}(t^{ba}_{kj})_{\text{MP$2$}} \ .
    \end{split}
\end{gather}
\end{subequations}
Here, we introduce the MP2 CCD amplitude notation as $(t^{ab}_{ij})_{\text{MP$2$}} = -\frac{\braket{ab||ij}}{\Delta^{ab}_{ij}}$, with $\Delta^{ab}_{ij}=\epsilon_{a}+\epsilon_{b}-\epsilon_{i}-\epsilon_{j}$ and $\braket{pq||rs}=\braket{pq|rs}-\braket{pq|sr}$ representing the anti-symmetrized two-electron repulsion integrals. The forward-time interaction matrices are given by 
\begin{subequations}
\begin{gather}
    \begin{split}~\label{eq:forward}
        (\mathbf{K}^{>,2\p1\h}_{iab,jcd}+\mathbf{C}^{>,2\p1\h}_{iab,jcd}) &= (\epsilon_{a}+\epsilon_{b}-\epsilon_{i})(\delta_{ac}\delta_{bd}-\delta_{ad}\delta_{bc})\delta_{ij}\\
        &+\braket{ab||cd}\delta_{ij}
        +\braket{jb||di}\delta_{ac}\\
        &- \braket{jb||ci}\delta_{ad}
    +\braket{ja||ci}\delta_{bd}\\
    &- \braket{ja||di}\delta_{bc}\ ,
    \end{split}
\end{gather}
with the backward-time interaction matrices as 
\begin{gather}
    \begin{split}~\label{eq:backward}
        (\mathbf{K}^{<,2\h1\p}_{ija,klb}+\mathbf{C}^{<,2\h1\p}_{ija,klb}) &= (\epsilon_{i}+\epsilon_{j}-\epsilon_{a})\delta_{ab}(\delta_{ik}\delta_{jl}-\delta_{il}\delta_{jk })\\
        &-\braket{ij||kl}\delta_{ab}-\braket{jb||al}\delta_{ik}\\
        &+ \braket{jb||ak}\delta_{il}
    +\braket{ib||al}\delta_{jk}\\
    &- \braket{ib||ak}\delta_{jl}\ .
    \end{split}
\end{gather} 
\end{subequations}
Using these expressions for the coupling and interaction matrices in the Dyson supermatrix representation (Eq.~\ref{eq:el_dys}) corresponds to an infinite-order summation of the electronic self-energy diagrams depicted in Fig.~\ref{fig:3rd_order}. To identify the connection to coupled-cluster doubles theory, we define $\Sigma^{\infty} =0 $, as the reference zeroth-order Hamiltonian is the Fock operator. 

\section{Relationship between the electronic self-energy and coupled-cluster doubles theory}

In the following we show that the connection between the electronic self-energy and the CCD amplitude equations can be derived by decoupling the forward and backward time self-energy components along with separating the particle-hole sectors. The decoupling of the different time directions and particle-hole sectors is commonly referred to as a non-Dyson self-energy approximation.~\cite{lange2018relation,bintrim2021full,ortiz2020dyson,hirata2024nonconvergence} 
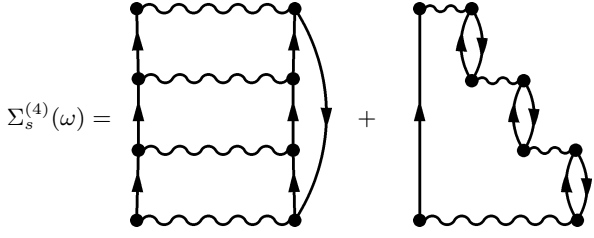
\begin{figure}[ht]
    \centering
    \begin{gather*}
\begin{split}~\label{eq:cc_III_pt}
\Sigma^{(4)}_{s}(\omega)&= \hspace{2.5mm}
    \begin{gathered}
    \begin{fmfgraph*}(60,80)
    \fmfcurved
    \fmfset{arrow_len}{3mm}
    \fmfleft{i1,i2,i3,i4}
    \fmflabel{}{i1}
    \fmflabel{}{i2}
    \fmfright{o1,o2,o3,o4}
    \fmflabel{}{o1}
    \fmflabel{}{o2}
    \fmf{wiggly}{i2,o2}
    \fmf{wiggly}{i3,o3}
    \fmf{wiggly}{i1,o1}
    \fmf{wiggly}{i4,o4}
    \fmf{fermion}{i1,i2}
    \fmf{fermion}{i2,i3}
    \fmf{fermion}{i3,i4}
    \fmf{fermion}{o1,o2}
    \fmf{fermion}{o2,o3}
    \fmf{fermion}{o3,o4}
    \fmf{fermion,left=0.3}{o4,o1}
    \fmfforce{(0.0w,0.0h)}{i1}
    \fmfforce{(1.0w,0.0h)}{o1}
    \fmfforce{(0.0w,1.0h)}{i4}
    \fmfforce{(1.0w,1.0h)}{o4}
    \fmfdot{i1,i2,i3,i4}
    \fmfdot{o1,o2,o3,o4}
\end{fmfgraph*}
\end{gathered}
\hspace{7.5mm}+\hspace{5mm}
\begin{gathered}
\begin{fmfgraph*}(60,80)
    \fmfcurved
    \fmfset{arrow_len}{3mm}
    \fmfleft{i1,i2,i3,i4}
    \fmflabel{}{i1}
    \fmflabel{}{i2}
    \fmfright{o1,o2,o3,o4}
    \fmflabel{}{o1}
    \fmflabel{}{o2}
    \fmf{wiggly}{i1,o1}
    \fmf{wiggly}{o2,v2}
    \fmf{wiggly}{v3,v4}
    \fmf{wiggly}{i4,o4}
    \fmf{fermion}{i1,i4}
    \fmf{fermion,left=0.3}{o1,o2}
    \fmf{fermion,left=0.3}{o2,o1}
    \fmf{fermion,left=0.3}{v4,o4}
    \fmf{fermion,left=0.3}{o4,v4}
    \fmf{fermion,left=0.3}{v2,v3}
    \fmf{fermion,left=0.3}{v3,v2}
    \fmfforce{(0.0w,0.0h)}{i1}
    \fmfforce{(1.0w,0.0h)}{o1}
    \fmfforce{(0.66w,0.33h)}{v2}
    \fmfforce{(0.33w,0.0h)}{v1}
    \fmfforce{(0.66w,0.66h)}{v3}
    \fmfforce{(0.33w,0.66h)}{v4}
    \fmfforce{(0.0w,1.0h)}{i4}
    \fmfforce{(0.33w,1.0h)}{o4}
    \fmfdot{v2,v3,v4}
    \fmfdot{i1,i4}
    \fmfdot{o1,o2,o4}
\end{fmfgraph*}
\end{gathered}
\end{split}
\end{gather*}
\caption{The fourth-order one-particle irreducible skeleton electronic self-energy diagrams contained in the full CCD equations when their contribution is included in the set of 2p1h/2h1p interaction matrices.}
    \label{fig:4th_order}
\end{figure}
The connections presented in this section are related to but distinct from the diagrammatic coupled-cluster self-energy, recently introduced in Refs~\citenum{coveney2023coupled} and~\citenum{coveney2024cc_se}, which is defined over the complete set of occupied and virtual orbitals. As a result, the non-Dyson electronic self-energy uncovered in this work that yields the exact CCD amplitude equations cannot be derived by taking functional derivatives of a diagrammatic expansion for the CC ground-state energy.

We refer to this non-Dyson electronic self-energy approximation as the \emph{particle-hole-time decoupled self-energy}. For the occupied states, we have
\begin{gather}
    \begin{split}~\label{eq:occ_se}
        \bar{\Sigma}^{\F}_{ij}(\omega) &= \sum_{\substack{ka>b\\lc>d}}\bar{U}_{i,kab}\\
        &\times\Big((\omega+i\eta)\mathbbm{1}-(\mathbf{K}^{>,2\p1\h}+\mathbf{C}^{>,2\p1\h})\Big)^{-1}_{kab,lcd}U_{lcd,j}
    \end{split}
\end{gather}
where all backward-time contributions are decoupled. This is motivated by the fact that the coupled-cluster excitation operators contain only hole to particle excitations and so are restricted to only contain forward-time propagation: $T_2=\sum_{\substack{b>a\\j>i}}t^{ab}_{ij}a^\dag_aa^\dag_ba_ja_i$. 
Additionally, we also neglect all forward-time self-energy diagrams that contain internal backward-time propagation of the intermediate Green's function lines from the final interaction vertex such that: $\bar{U}_{i,jab} = \braket{ij||ab}$. This procedures ensures that the correct set of only forward-time ordered Feynman-Goldstone self-energy diagrams are included in the particle-hole-time decoupled self-energy. However, the form of the second coupling matrix, $U_{lcd,j}$ remains the same as defined in Eq.~\ref{eq:int_forward}. In general, this breaks the hermiticity of the coupling elements of the electronic self-energy. This retention of specific forward-time ordered diagrams is closely related to well-known approximations for the polarization propagator. The full set of time-orderings contained in the ADC(3) supermatrix contain both forward- and backward-time bubble, exchange and ladder polarization insertions. This is also the case in the $GW$-RPA approximation whereby both forward- and backward-time bubble diagrams are present.~\cite{berkelbach2018communication,lange2018relation} The inclusion of both time-orderings of these polarization diagrams is particularly important for both electronic relaxation and correlation processes.

By particle-hole symmetry, we have the particle-hole-time decoupled self-energy of the virtual states as 
\begin{gather}
    \begin{split}~\label{eq:vir_se}
        \bar{\Sigma}^{\B}_{ab}(\omega) &= \sum_{\substack{i>jc\\k>ld}}\bar{V}_{a,ijc}\\
        &\times\Big((\omega-i\eta)\mathbbm{1}-(\mathbf{K}^{<,2\h1\p}+\mathbf{C}^{<,2\h1\p})\Big)^{-1}_{ijc,kld}V^\dag_{kld,b} \ ,
    \end{split}
\end{gather}
where now we decouple all forward-time contributions and neglect all diagrams that contain internal forward-time propagation of the intermediate Green's function lines from the final interaction verex such that: $\bar{V}_{a,ijc} = \braket{ac||ij}$. The second coupling matrix element, $V^\dag_{kld,b}$, remains the same as defined in Eq.~\ref{eq:int_backward}. Decoupling the particle-hole sectors requires: $\bar{\Sigma}_{ia} = \bar{\Sigma}_{ai} = 0$. 

Focusing on the particle-hole-time decoupled self-energy of the occupied-occupied block, Eq.~\ref{eq:occ_se}, we have the upfolded supermatrix eigenvalue problem
\begin{gather}
    \begin{split}~\label{eq:up}
        \left(\begin{array}{cc}
            \epsilon_{i}\delta_{ij} & \bar{U}_{i,jab} \\
            U_{kcd,i} & (\mathbf{K}^{>,2\p1\h}_{iab,jcd}+\mathbf{C}^{>,2\p1\h}_{iab,jcd}) 
        \end{array}\right)\left(\begin{array}{c}
             X^{\nu}_{i}  \\
              Y^{\nu}_{iab}
        \end{array}\right) = \left(\begin{array}{c}
             X^{\nu}_{i}  \\
              Y^{\nu}_{iab}
        \end{array}\right) \varepsilon_{\nu} \ .
    \end{split}
\end{gather}
Now, we introduce the doubles amplitudes through the follow identity: $t^{ab}_{ij}=\sum_{\nu}Y^{\nu}_{jab}(X^{-1})^{\nu}_{i}$. In matrix notation, we write $(\mathbf{T})^{ab}_{ij}=(\mathbf{Y}\mathbf{X}^{-1})^{ab}_{ij}=t^{ab}_{ij}$. Using this identity with Eq.~\ref{eq:up}, we have
\begin{gather}
    \begin{split}~\label{eq:ccd_super}
         \left(\begin{array}{cc}
            \mathbf{f} & \mathbf{\bar{U}} \\
            \mathbf{U} & \mathbf{K}^{>,2\p1\h}+\mathbf{C}^{>,2\p1\h} 
        \end{array}\right)\left(\begin{array}{c}
             \mathbbm{1} \\
              \mathbf{T}
        \end{array}\right) = \left(\begin{array}{c}
             \mathbbm{1} \\
              \mathbf{T}
        \end{array}\right)\mathbf{X}\mathbf{\mathcal{E}}\mathbf{X}^{-1} \ .
    \end{split}
\end{gather}
These simultaneous equations give the extended Fock operator and the self-energy Riccati equation:
\begin{subequations}
    \begin{align}
    \begin{split}
        \mathbf{F}\mathbf{X} &= \mathbf{X}\mathbf{\mathcal{E}}
        \end{split}\\
    \begin{split}~\label{eq:se_ricc}
        \mathbf{U} + (\mathbf{K}^{>,2\p1\h}+\mathbf{C}^{>,2\p1\h})\mathbf{T} -\mathbf{T}\mathbf{F}  &= \mathbf{0} \ ,
    \end{split}
\end{align}
\end{subequations}
where $\mathbf{F} = \mathbf{f} + \mathbf{\bar{U}}\mathbf{T}$ is the extended Fock operator. In explicit matrix notation, the extended Fock operator is written as
\begin{gather}
    \begin{split}~\label{eq:ext_fock}
        F_{ij} = \epsilon_{i}\delta_{ij} + \frac{1}{2}\sum_{kab}\braket{ik||ab}t^{ab}_{jk} \ .
    \end{split}
\end{gather}
The extended Fock operator is exactly of the form of the upper left block in IP-EOM-CCD theory: $F_{ij}=-\braket{\Phi_{j}|\bar{H}_N|\Phi_{i}}$, where $\bar{H}_N = e^{-T_2}He^{T_2}-E^{\CC}_0$ is the normal-ordered CCD similarity transformed Hamiltonian.~\cite{scuseria1995connections,coveney2023coupled,coveney2024cc_se} Eq.~\ref{eq:ext_fock} is also found by neglecting the $T_1$ amplitudes appearing in Refs~\citenum{gauss1995coupled,nooijen1995equation,nooijen1997similarity,musial2003equation}. Using the coupling and interaction matrix elements defined above in Eqs~\ref{eq:int_forward} and~\ref{eq:forward}, the self-energy Riccati equation is explicitly written as 
\begin{gather}
    \begin{split}~\label{eq:ADC_CCD}
        &\braket{ab||ij} +\frac{1}{2}\sum_{kl}(t^{ab}_{kl})_{\text{MP$2$}}\braket{ij||kl} - \sum_{kc}\braket{kb||ic}(t^{ac}_{kj})_{\text{MP$2$}}\\
        &+\sum_{kc}\braket{ka||ic}(t^{bc}_{kj})_{\text{MP$2$}} +\Delta^{ab}_{ij}t^{ab}_{ij} + \frac{1}{2}\sum_{cd}\braket{ab||cd}t^{cd}_{ij}\\
        &-\sum_{kc}\braket{kb||jc}t^{ac}_{ik}+\sum_{kc}\braket{ka||jc}t^{bc}_{ik}\\
        &-\frac{1}{2}\sum_{klcd}t^{cd}_{ik}\braket{kl||cd}t^{ab}_{jl} = 0 \ .
    \end{split}
\end{gather}
The self-energy Riccati equation as defined in Eq~\ref{eq:se_ricc} represents a subset of the full CCD equations~\cite{shavitt2009many}:
\begin{gather}
    \begin{split}~\label{eq:full_CCD}
        &\braket{ab||ij} +\Delta^{ab}_{ij}t^{ab}_{ij} + \frac{1}{2}\sum_{cd}\braket{ab||cd}t^{cd}_{ij} +\frac{1}{2}\sum_{kl}\braket{ij||kl}t^{ab}_{kl}\\
        &-\sum_{kc}(\braket{kb||jc}t^{ac}_{ik}
        -\braket{ka||jc}t^{bc}_{ik}+\braket{ak||ci}t^{bc}_{jk}-\braket{bk||ci}t^{ac}_{jk})\\
        &\sum_{klcd}\braket{kl||cd}\Big(\frac{1}{4}t^{ab}_{kl}t^{cd}_{ij}-\frac{1}{2}(t^{ab}_{ik}t^{cd}_{jl}+t^{cd}_{ik}t^{ab}_{jl} +t^{ac}_{ij}t^{bd}_{kl}+t^{bd}_{ij}t^{ac}_{kl})\\
        &+(t^{ac}_{ik}t^{bd}_{jl}+t^{bd}_{ik}t^{ac}_{jl})\Big)= 0 \ .
    \end{split}
\end{gather}
As can be seen by analysis of Eq.~\ref{eq:full_CCD}, the self-energy CCD amplitude equations given in Eq.~\ref{eq:ADC_CCD} are missing several additional terms. These missing contributions include the particle-particle ladder terms, the full antisymmetrized ring diagrams as well as the six additional quadratic terms. Another difference is that only the MP2 doubles amplitudes appear in the hole-hole ladder and ring terms of the self-energy amplitude equations. These effectively represent a first iteration of the CCD equations using the MP2 amplitudes as an initial guess.

Our approach shares several similarities to the rCCD amplitude equations, first derived in Ref.~\citenum{scuseria2008ground}. However, the rCCD amplitude equations are given by~\cite{scuseria2008ground,scuseria2013particle,rishi2020route,ring2004nuclear,berkelbach2018communication} 
\begin{gather}
    \begin{split}
        &\braket{ab||ij} +\Delta^{ab}_{ij}t^{ab}_{ij}-\sum_{kc}\braket{kb||jc}t^{ac}_{ik}\\
        &-\sum_{kc}\braket{ka||ic}t^{bc}_{jk}+\sum_{klcd}t^{ac}_{ik}\braket{kl||cd}t^{bd}_{jl} = 0 \ ,
    \end{split}
\end{gather}
and originate from the RPA eigenvalue problem. The RPA approximation only contains ring diagram contributions to the CCD equations that are not antisymmetrized with respect to exchange of the external indices. 

It should be noted that the MP2 self-energy approximation corresponds to taking the interaction matrices to be: $\mathbf{K}^{>,\text{MP}2}_{iab,jcd}=(\epsilon_{a}+\epsilon_{b}-\epsilon_{i})(\delta_{ac}\delta_{bd}-\delta_{ad}\delta_{bc})\delta_{ij}$. This yields the effective doubles amplitude equations for the particle-hole-time decoupled MP2 self-energy:
\begin{gather}
    \begin{split}
        \braket{ab||ij} +\Delta^{ab}_{ij}t^{ab}_{ij} -\frac{1}{2}\sum_{klcd}t^{cd}_{ik}\braket{kl||cd}t^{ab}_{jl} = 0 \ .
    \end{split}
\end{gather}

We now demonstrate how to obtain the full CCD equations from the electronic self-energy. To do so we must transform the coupling and interaction matrices to become self-consistently dependent on the corresponding amplitude solutions. For the coupling matrices, Eq.~\ref{eq:int_forward}, this simply corresponds to replacing all MP2 amplitudes with the exact doubles amplitudes to be determined: $(t^{ab}_{ij})_{\text{MP2}}\rightarrow t^{ab}_{ij}$. This results in the `self-consistent' coupling matrix elements:
\begin{gather}
    \begin{split}~\label{eq:sc_ints}
        U^{\text{sc}}_{abj,p}&=\braket{ab||pj} +\frac{1}{2}\sum_{kl}t^{ab}_{kl}\braket{pj||kl} - \sum_{kc}\braket{kb||pc}t^{ac}_{kj}\\
        &+\sum_{kc}\braket{ka||pc}t^{bc}_{kj} \ .
    \end{split}
\end{gather}
These coupling matrix elements can be viewed in terms of a self-consistent Green's function theory whereby the solution is updated iteratively. 
For the interaction matrices, the following self-consistent updates are necessary:
\begin{subequations}~\label{eq:mod_ints}
    \begin{align}
    \begin{split}
        \epsilon_i\delta_{ij} &\rightarrow \epsilon_i\delta_{ij} +\frac{1}{2}\sum_{kcd} \braket{ik||cd}t^{cd}_{jk}
    \end{split}\\
    \begin{split}
        \epsilon_a\delta_{ab} &\rightarrow \epsilon_a\delta_{ab} -\frac{1}{2}\sum_{ijc} \braket{ij||bc}t^{ca}_{ij}
        \end{split}\\
    \begin{split}
        \braket{ab||cd} &\rightarrow \braket{ab||cd} +\frac{1}{2}\sum_{kl}\braket{kl||cd} t^{ab}_{kl} 
         \end{split}\\
    \begin{split}
            \braket{ja||bi} &\rightarrow \braket{ja||bi} + \sum_{kc}  \braket{jk||bc} t^{ac}_{ik} \ .
    \end{split}
\end{align}
\end{subequations}
Including these additional terms in the interaction matrices, $(\mathbf{K}^{>,2\p1\h}_{iab,jcd}+\mathbf{C}^{>,2\p1\h}_{iab,jcd})$, (see Eq.~\ref{eq:forward}) that enter the self-energy Riccati equation (Eq.~\ref{eq:se_ricc}) gives rise to the six additional quadratic terms required to generate the full CCD equations given in Eq.~\ref{eq:full_CCD}. The additional terms, generated by the replacements given in Eqs~\ref{eq:mod_ints}, can be viewed as self-consistent fourth-order electronic self-energy diagrams that contain contributions in the 2p1h/2h1p excitation space (see Fig.~\ref{fig:4th_order} and Appendix~\ref{app:2}) that are present in the ADC(4) construction.~\cite{schirmer1983new} However, in the perturbative ADC(4) construction their contribution is restricted to contain MP2 doubles amplitudes only and necessarily requires the expansion of the Dyson supermatrix to include 3p2h/3h2p intermediate state configurations.~\cite{schirmer1983new,schirmer2018many} The self-energy Riccati equation, Eq.~\ref{eq:se_ricc}, resulting from the replacements made in Eqs~\ref{eq:sc_ints} and~\ref{eq:mod_ints} to the coupling and interaction matrices gives the full CCD amplitude equations. These results demonstrate that the full CCD equations can be viewed as an infinite partial summation through fourth-order of perturbation theory in the electronic self-energy when restricted to 2p1h/2h1p excitation character of the Intermediate State Configurations. The Feynman-Goldstone diagrams for the particle-hole-time decoupled self-energy that generate the full CCD amplitude equations are given in Appendix~\ref{app:2}. To the best of our knowledge, this is the first time that the \emph{complete} CCD amplitude equations have been derived within the Green's function formalism. 

It is well established that the singles amplitudes, $\{t^{a}_{i}\}$, can be formally eliminated by the use of Brueckner orbitals whereby the spin-orbitals are determined simultaneously with the optimization of the cluster amplitudes.~\cite{chiles1981electron,handy1989size,scuseria1995connections,helgaker2013molecular,tew2016explicitly} This orbital rotation determined by the singles amplitude equations is termed Brueckner coupled-cluster theory (BCC). Therefore, upon rotation to the Brueckner orbitals the derivation of the doubles amplitude equations presented above is unchanged. Additionally, the formalism presented in this work can also be extended to derive the higher body BCCDT amplitude equations by retention and identification of the correct time-ordered self-energy diagrams contained in the higher-order ADC$(n)$ approximations.

The ground state correlation energy is found from the relation
\begin{gather}
    \begin{split}~\label{eq:gs_energy}
        E^N_0 &= \frac{1}{2}\tr\mathbf{X}\mathbf{\mathcal{E}}\mathbf{X}^{-1} +\frac{1}{2}\tr[\mathbf{h}]\\
        &= E^{\HF}_0 + \frac{1}{2}\tr[\mathbf{\bar{U}T}]\\
        &= E^{\HF}_0 + \frac{1}{4}\sum_{ijab}\braket{ij||ab}t^{ab}_{ij} \ ,
    \end{split}
\end{gather}
where we have used the identity: $\mathbf{X}\mathbf{\mathcal{E}}\mathbf{X}^{-1}=\mathbf{F}$ from Eq.~\ref{eq:ext_fock} to go from the first line to second line of Eq.~\ref{eq:gs_energy}. This expression is exactly equivalent to that obtained from CCD theory. Eq.~\ref{eq:gs_energy} demonstrates that we can obtain correlation energies from Green's function theory that can be cast exactly in terms of CCD theory.

The doubles amplitude equations generated by the particle-hole-time decoupled self-energy for the virtual block can be found in Appendix~\ref{app:1} and are identical through particle-hole symmetry.

The relationship to the IP-EOM-CCD eigenvalue problem can be seen by coupling the self-consistent backward-time particle-hole-time decoupled self-energy (see Eq.~\ref{eq:vir_se} and Appendix~\ref{app:1}) to the occupied-occupied block of the self-energy while maintaining the particle-hole separation. This coupling is necessary to account for the negation of the full time-orderings of the coupling matrices of the forward-time particle-hole-time decoupled self-energy (see Eq.~\ref{eq:occ_se}) in order to obtain accurate approximations for the removal excitation energies. Inclusion of these terms results in equations that are analogous to those of IP-EOM-CCD theory.~\cite{nooijen1995equation,nooijen1997similarity,stanton1999simple,hirata2000high,hirata2000high1,musial2003equation} As particle-hole separation is maintained, the resulting self-energy approximation is still of the non-Dyson form. Therefore, the presence of the lambda de-excitation amplitudes, $\{\lambda_{\mu}\}$ required to construct the left eigenstate of $\bar{H}$, that appear in Refs~\citenum{coveney2023coupled} and~\citenum{coveney2024cc_se} in the diagrammatic coupled-cluster self-energy do not appear in this work. As the non-Dyson approximation is kept, the connection can be explicitly made to IP/EA-EOM-CCD theory, where the particle-hole sectors are decoupled. The coupling of backward-time components to the particle-hole-time decoupled self-energy of the occupied-occupied block gives the additional particle-hole decoupled self-energy contribution 
\begin{gather}
    \begin{split}
\bar{\Sigma}^{\B}_{ij}(\omega) &= \sum_{\substack{k>la\\m>nb}}V^{\text{sc}}_{i,kla}\\
        &\times\Big((\omega-i\eta)\mathbbm{1}-(\mathbf{\bar{K}}^{<,2\h1\p}+\mathbf{\bar{C}}^{<,2\h1\p})\Big)^{-1}_{kla,mnb}\bar{V}^\dag_{mnb,j} ,
        \end{split}
\end{gather}
where we define the transformed coupling elements from the self-consistent coupling matrices given in Appendix~\ref{app:1} and reproduced here for clarity:
\begin{subequations}
\begin{gather}
    \begin{split}
        V^{\text{sc}}_{i,kla} &=  \braket{ia||kl} + \sum_{cm}\braket{im||kc}t^{ca}_{ml} -\sum_{cm}\braket{im||lc}t^{ca}_{mk} \\
        &+\frac{1}{2}\sum_{cd} \braket{ia||cd}t^{cd}_{kl}
        \end{split}
        \end{gather}
        with
        \begin{gather}
    \begin{split}
         \bar{V}^\dag_{mnb,j} &= \braket{mn||jb} \ .
    \end{split}
\end{gather}
\end{subequations}
These coupling matrix elements are exactly equal to the blocks of the CCD similarity transformed Hamiltonian: $V_{i,kla} = \braket{\Phi^{a}_{kl}|\bar{H}_N|\Phi_{i}}$ and $\bar{V}^\dag_{mnb,j} = \braket{\Phi_{j}|\bar{H}_N|\Phi^{b}_{mn}}$.~\cite{shavitt2009many,lange2018relation}

The self-consistent interaction matrices are exactly those appearing in Appendix~\ref{app:1} and are written compactly as~\cite{coveney2023coupled,coveney2024cc_se}
\begin{gather}
    \begin{split}~\label{eq:interactions}
        (\mathbf{\bar{K}}^{<,2\h1\p}_{ija,klb}+\mathbf{\bar{C}}^{<,2\h1\p}_{ija,klb}) &= F_{ik}\delta_{ab}\delta_{jl}+ F_{jl}\delta_{ab}\delta_{ik}-F_{ba}\delta_{ik}\delta_{jl} \\
        &- F_{il}\delta_{ab}\delta_{jk}- F_{jk}\delta_{ab}\delta_{il}+F_{ba}\delta_{il}\delta_{jk}\\&
        +\chi_{ib,al}\delta_{jk}+\chi_{jb,ak}\delta_{il}-\chi_{ij,kl}\delta_{ab} \\
    &-\chi_{ib,ak}\delta_{jl}- \chi_{jb,al}\delta_{ik} \ .
    \end{split}
\end{gather}
We maintain particle-hole separability such that $\bar{\Sigma}_{ia}=\bar{\Sigma}_{ai}=0$. The interaction matrices are almost identical to the matrix elements of the CCD similarity transformed Hamiltonian in the basis of 2h1p determinants: $(\mathbf{\bar{K}}^{<,2\h1\p}_{ija,klb}+\mathbf{\bar{C}}^{<,2\h1\p}_{ija,klb})\approx-\braket{\Phi^{b}_{kl}|\bar{H}_N|\Phi^{a}_{ij}}$. However, the explicit three-body interaction, $\chi_{ijb,kal}$, that arises in IP-EOM-CCD theory is not present in Eq.~\ref{eq:interactions}.~\cite{coveney2023coupled,coveney2024cc_se,nooijen1995equation,nooijen1997similarity,musial2003equation,hirata2004higher}
The result of this coupling is the effective particle-hole decoupled quasiparticle Hamiltonian
\begin{gather}
    \begin{split}
        \hat{H}&^{\text{IP-EOM-CCD}}_{ij}(\omega) = F_{ij} + \bar{\Sigma}^{\B}_{ij}(\omega) \ ,
    \end{split}
\end{gather}
from which we have the upfolded supermatrix representation:
\begin{gather}
    \begin{split}~\label{eq:up_eom}
        \mathbf{\bar{D}}^{\text{IP-EOM-CCD}}=\left(\begin{array}{cc}
            \mathbf{f}+\mathbf{\bar{U}T} & \mathbf{V}^{\text{sc}} \\
            \mathbf{\bar{V}}^\dag & \mathbf{\bar{K}}^{<,2\h1\p}+\mathbf{\bar{C}}^{<,2\h1\p} 
        \end{array}\right) . 
    \end{split}
\end{gather}
Explicitly, we can write this supermatrix in the suggestive form
\begin{gather}
    \begin{split}~\label{eq:eom_conn}
        \mathbf{\bar{D}}^{\text{IP-EOM-CCD}}=\left(\begin{array}{cc}
            -\braket{\Phi_{j}|\bar{H}_N|\Phi_{i}} &  \braket{\Phi^{a}_{kl}|\bar{H}_N|\Phi_{i}} \\
           \braket{\Phi_{j}|\bar{H}_N|\Phi^{b}_{mn}}  & -\braket{\Phi^{b}_{mn}|\bar{H}_N|\Phi^{a}_{kl}} 
        \end{array}\right) ,
    \end{split}
\end{gather}
remembering that the interaction matrices (bottom right block of the supermatrix) do not include the three-body CC interaction. Supermatrices of the form of Eqs~\ref{eq:up_eom}/\ref{eq:eom_conn} have structures that are analogous to the IP-EOM-CCD supermatrix as demonstrated in Refs~\citenum{coveney2023coupled} and~\citenum{coveney2024cc_se}. The eigenvalues of $\mathbf{\bar{D}}^{\text{IP-EOM-CCD}}$ provide access to approximate electron removal energies. However, the residues of the single-particle Green's function are not obtained within this approximation as the particle-hole sectors remain decoupled. The relationship to EA-EOM-CCD theory can be derived in an entirely analogous manner by coupling the forward time self-energy contributions to the virtual states (see Eq.~\ref{eq:vir_se}) while maintaining particle-hole separability.

\section{Application to a model system: Exact solution of the Hubbard Dimer}

We apply the formalism presented in this work to the exactly solvable Hubbard dimer at half filling. The Hubbard model is parametrized by the ratio $\frac{U}{t}$, which provides a measure of the `correlation strength'. The parameters $U$ and $t$ represent the on-site repulsion and the coupling strength between nearest neighbour sites, respectively.~\cite{hubbard1964electron} The CCD amplitude equations are exact for the Hubbard dimer and their quadratic Riccati equation yields two solutions: 
\begin{equation}~\label{eq:exact_CCD}
(t^{aa}_{ii})_{\pm} = \frac{4t\pm c}{U} \ ,
\end{equation}
where $c=\sqrt{U^2+16t^2}$. The CCD amplitude is the negative solution, $(t^{aa}_{ii})_{-}$, which gives the exact ground state correlation energy: $E_{c} =\frac{U}{2}(t^{aa}_{ii})_{-}= 2t-\frac{c}{2}$.

In Ref.~\citenum{coveney2023regularized}, we demonstrated that the second-order MP2 self-energy is exact for the Hubbard dimer. From the MP2 self-energy, the corresponding Dyson supermatrix for this system splits into the following eigenvalue problems for the occupied and virtual blocks respectively 
\begin{subequations}
\begin{align}
	\begin{split}~\label{eq:mp2_forward}
		\left(\begin{array}{cc}
			\epsilon_i & \frac{U}{2}\\
			\frac{U}{2} & 2\epsilon_a-\epsilon_i \\ 
		\end{array}\right)\left(\begin{array}{c}
		X \\
		Y
		\end{array}\right) &= \left(\begin{array}{c}
		X \\
		Y
		\end{array}\right) \mathbf{\varepsilon}^{\h}
        \end{split}\\
        \begin{split}
    		\left(\begin{array}{cc}
			 \epsilon_a & \frac{U}{2}\\
			\frac{U}{2} & 2\epsilon_i-\epsilon_a \\ 
		\end{array}\right)\left(\begin{array}{c}
			\tilde{X} \\
			\tilde{Y}
		\end{array}\right) &= \left(\begin{array}{c}
			\tilde{X} \\
			\tilde{Y}
		\end{array}\right) \mathbf{\varepsilon}^{\p}
		\end{split}
\end{align}
\end{subequations}
where $\epsilon_{i/a}=\frac{U}{2}\mp t$ are the HF reference energies of the occupied/virtual states, respectively. 

Multiplying Eq.~\ref{eq:mp2_forward} by $X^{-1}$, we obtain the extended Fock operator and self-energy Riccati equation as
\begin{subequations}
    \begin{align}
        \begin{split}~\label{eq:mp2_fock}
            \epsilon_i + \frac{U}{2}t^{aa}_{ii} &= \varepsilon^{\h}
        \end{split}\\
        \begin{split}~\label{eq:ricc}
            \frac{U}{2} +2(\epsilon_a-\epsilon_i)t^{aa}_{ii} - \frac{U}{2}(t^{aa}_{ii})^2 &= 0 \ ,
        \end{split}
    \end{align}
\end{subequations}
\begin{figure}[ht]
    \centering
    \includegraphics[width=90mm,height=80mm]{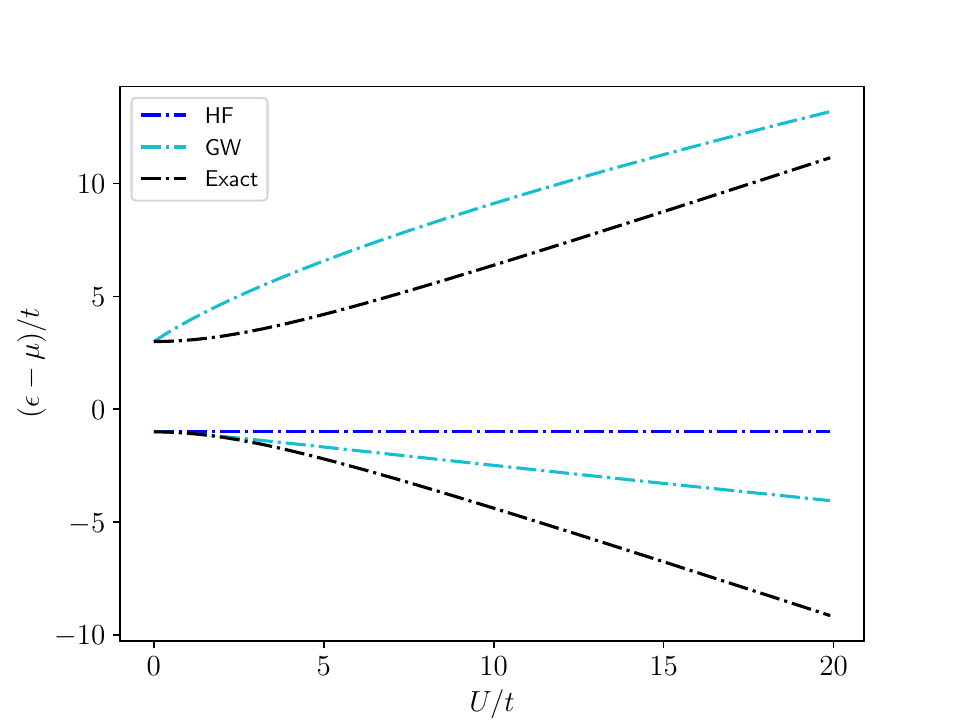}
    \caption{Hole quasiparticle and satellite energies of the Hubbard dimer as a function of $\frac{U}{t}$ relative to the chemical potential, $\mu=\frac{U}{2}$. The particle states are related to the hole states by particle-hole symmetry.}
    \label{fig:qps}
\end{figure}
where we have used the identity: $t^{aa}_{ii}=Y\cdot X^{-1}$. The self-energy Riccati equation, Eq.~\ref{eq:ricc}, is the exact CCD amplitude equation:
\begin{gather}
    \begin{split}
        \frac{U}{2}(t^{aa}_{ii})^2 - 4t(t^{aa}_{ii}) - \frac{U}{2} = 0 \ , 
    \end{split}
\end{gather}
and yields two solutions: $(t^{aa}_{ii})_{\pm}=\frac{1}{U}(4t\pm c)$. The corresponding hole eigenvalues and super-eigenvectors of the quasiparticle and satellite solutions are found by substitution of the CCD amplitude solutions,  $(t^{aa}_{ii})_{\pm}$, into Eq.~\ref{eq:mp2_fock} to give
\begin{subequations}~\label{eq:hole}
    \begin{align}
	\begin{split}
    \varepsilon^{\h}_{\text{QP}} &= \epsilon_i + \frac{U}{2}(t^{aa}_{ii})_{-} \hspace{2.5mm};\hspace{2.5mm}
		\mathbf{v}^{\h}_{\text{QP}} = \left(\begin{array}{c}
			1 \\
			(t^{aa}_{ii})_{-}
		\end{array}\right)  
	\end{split}\\
	\begin{split}
    \varepsilon^{\h}_{\sat} &= \epsilon_i + \frac{U}{2}(t^{aa}_{ii})_{+} \hspace{2.5mm};\hspace{2.5mm}
		\mathbf{v}^{\h}_{\sat} = \left(\begin{array}{c}
			1 \\
			(t^{aa}_{ii})_{+}
		\end{array}\right)  \ .
	\end{split}
\end{align}
\end{subequations}
The CCD amplitude is the lower component of the $\mathbf{v}^{\h}_{\text{QP}}$ vector of the hole quasiparticle solution: $(t^{aa}_{ii})_{-} = \frac{4t-c}{U}$. From Eq.~\ref{eq:renorm}, the inverse of the norm of the vector is exactly the hole quasiparticle renormalization factor: $Z^{\QP}_{h} = \left(\mathbf{v}^{\h\dag}_{\text{QP}} \mathbf{v}^{\h}_{\text{QP}}\right)^{-1}  =\Big(1+(t^{aa}_{ii})_{-}^2\Big)^{-1}=\frac{1}{2}+\frac{2t}{c}$. The hole satellite solution contains a lower component of the eigenvector, $\mathbf{v}^{\h}_{\sat}$, that is exactly the positive CCD amplitude solution: $(t^{aa}_{ii})_{+}=\frac{1}{U}(4t+c)$, with the satellite renormalization factor as $Z^{\sat}_{h} = \Big(1+(t^{aa}_{ii})_{+}^2\Big)^{-1}=\frac{1}{2}-\frac{2t}{c}$. As expected, the renormalization factor for the quasiparticle is larger than that of the satellite, $Z^{\QP}_{h}>Z^{\sat}_{h}$. Using Eq.~\ref{eq:gs_energy} for the ground state correlation energy gives
the exact ground state correlation energy of the Hubbard dimer.~\cite{romaniello2009self,romaniello2012beyond,lani2012approximations,martin2016interacting}

\begin{figure}[ht]
    \centering
    \includegraphics[width=90mm,height=80mm]{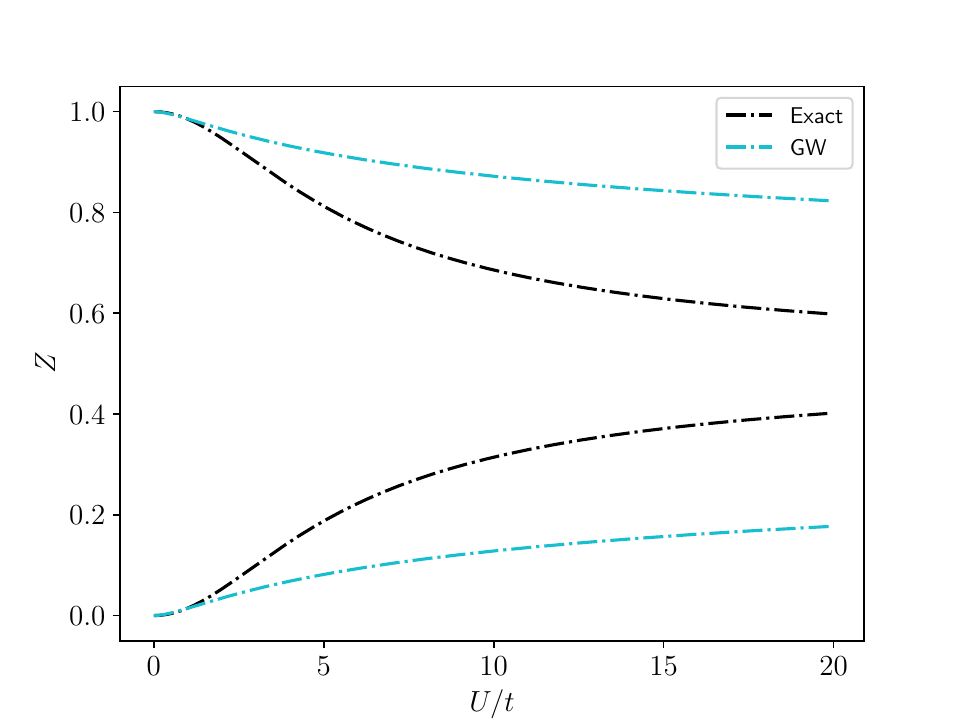}
    \caption{$Z^{\QP/\sat}_{\p/\h}$ renormalization factors for the Hubbard dimer as a function of $\frac{U}{t}$.}
    \label{fig:sat}
\end{figure}
For the particle solutions, the amplitude equations are directly connected to the hole solutions via particle-hole symmetry (see Appendix~\ref{app:1}). The corresponding particle eigenvalues and super-eigenvectors of the quasiparticle and satellite solutions are given by
\begin{subequations}~\label{eq:part}
    \begin{align}
	\begin{split}
    \varepsilon^{\p}_{\text{QP}} &= \epsilon_a- \frac{U}{2}(t^{aa}_{ii})_{-} \hspace{2.5mm};\hspace{2.5mm}
		\mathbf{v}^{\p}_{\text{QP}} = \left(\begin{array}{c}
			1 \\
			-(t^{aa}_{ii})^*_{-}
		\end{array}\right) 
	\end{split}\\
	\begin{split}
    \varepsilon^{\p}_{\sat} &= \epsilon_a - \frac{U}{2}(t^{aa}_{ii})_{+}\hspace{2.5mm};\hspace{2.5mm}
		\mathbf{v}^{\p}_{\sat} = \left(\begin{array}{c}
			1 \\
			-(t^{aa}_{ii})^*_{+}
		\end{array}\right) \ ,
	\end{split}
\end{align}
\end{subequations}
where we have used the relation from Appendix~\ref{app:1}: $\tilde{t}^{aa}_{ii} = \tilde{Y}\cdot\tilde{X}^{-1} = -(t^{aa}_{ii})^*$. The quasiparticle renormalization factor is again $Z^{\QP}_{p} = \Big(1+(t^{aa}_{ii})^2_{-}\Big)^{-1}$.
The satellite particle solution gives the lower component of the $\mathbf{v}^{\p}_{\sat}$ vector as the negative of the positive CCD solution: $-(t^{ab}_{ij})_{+}$, with the renormalization factor given by $Z^{\sat}_{p} = \Big(1+(t^{aa}_{ii})^2_{+}\Big)^{-1}$. Again, the renormalization factor of the quasiparticle is larger than that of the satellite, $Z^{\QP}_{p}>Z^{\sat}_{p}$. Using these expressions is it simple to verify that $Z^{\QP}_{h/p}+Z^{\sat}_{h/p}=1$, as required  by the sum rule.

\begin{figure}[ht]
    \centering
    \includegraphics[width=90mm,height=80mm]{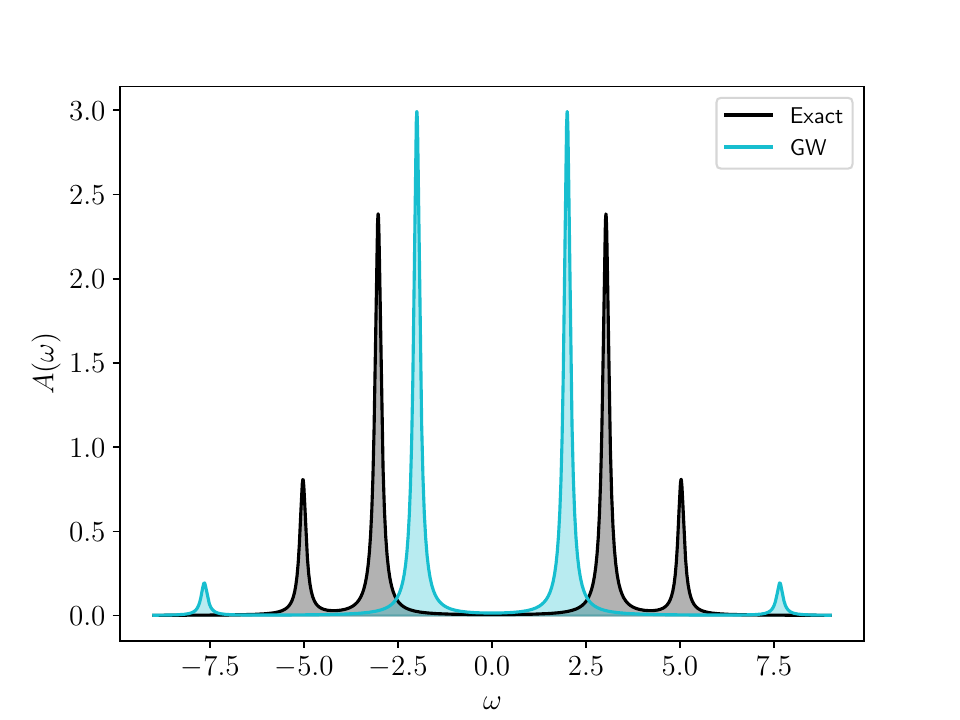}
    \caption{Exact and $GW$ spectral functions obtained for the Hubbard dimer at $\frac{U}{t}=7$. The exact spectral function is obtained from Eq.~\ref{eq:gf} as $A(\omega)=\frac{1}{\pi}\sum_{p}|\text{Im} G_{pp}(\omega)|$.}
    \label{fig:spectral}
\end{figure}
Substituting the expression for the CCD amplitudes (Eq.~\ref{eq:exact_CCD}) obtained from the self-energy Riccati equation (Eq.~\ref{eq:ricc}) into Eqs~\ref{eq:hole} and~\ref{eq:part}, we see that the particle and hole quasiparticle/satellite energies and renormalization factors are exact for the Hubbard dimer.~\cite{coveney2023regularized,riva2023multichannel,riva2024derivation,riva2025multichannel} The satellite and quasiparticle energies are exactly of the form of the extended Fock operators defined in Eqs~\ref{eq:ext_fock} and~\ref{eq:vir_fock}. These results concretely demonstrate that different CCD amplitude solutions can correspond to satellite excited states resulting from particle removal and addition processes. From our formulation, we can explicitly write the exact single-particle Green's function in terms of the CCD amplitudes:
\begin{subequations}~\label{eq:gf}
\begin{align}
    \begin{split}
        G_{ii}(\omega) &= \Big(1+(t^{aa}_{ii})^2_{-}\Big)^{-1}\frac{1}{\omega-(\epsilon_i+\frac{U}{2}(t^{aa}_{ii})_{-})-i\eta}\\
        &+\Big(1+(t^{aa}_{ii})^2_{+}\Big)^{-1}\frac{1}{\omega-(\epsilon_i+\frac{U}{2}(t^{aa}_{ii})_{+})+i\eta}
    \end{split}\\
    \begin{split}
        G_{aa}(\omega) &=\Big(1+(t^{aa}_{ii})^2_{-}\Big)^{-1}\frac{1}{\omega-(\epsilon_a-\frac{U}{2}(t^{aa}_{ii})_{-})+i\eta}\\
        &+\Big(1+(t^{aa}_{ii})^2_{+}\Big)^{-1}\frac{1}{\omega-(\epsilon_a-\frac{U}{2}(t^{aa}_{ii})_{+})-i\eta} \ .
    \end{split}
\end{align}
\end{subequations}
In Figs~\ref{fig:qps} and~\ref{fig:sat}, we plot the poles and weights of the exact Green's function given in Eq.~\ref{eq:gf} against the HF and $GW$ approximations obtained from Ref.~\citenum{coveney2023regularized}. In Fig.~\ref{fig:qps}, we see that the quasiparticle $GW$ solutions are close to the exact quasiparticle solutions obtained within our formalism for values of $\frac{U}{t}$ up to 3. Beyond this point, the quasiparticle $GW$ solutions begin to diverge away from the exact results as the essential correlated physics of electron localization is not recovered. In Fig.~\ref{fig:qps}, we also see that the $GW$ satellite solutions are completely qualitatively wrong. This is a result of the fact that the $GW$ approximation cannot handle the increasing multireference character of the underlying ground and excited state wavefunctions as the correlation strength $\frac{U}{t}$ increases.~\cite{coveney2023regularized,ammar2024can}

In Fig.~\ref{fig:sat}, we plot the quasiparticle and satellite renormalization factors as a function of $\frac{U}{t}$. We find that the $GW$ solutions diverge quickly from the exact results even in the weakly correlated regime as a result of the strong multireference character of the ground and excited state wavefunctions.~\cite{coveney2023regularized} In Fig.~\ref{fig:spectral}, we show the exact and $GW$ spectral functions, evaluated at $\frac{U}{t}=7$, obtained from the trace of the imaginary part of the corresponding Green's function. From analysis of the spectral function, we clearly see the qualitatively incorrect satellite peaks and strengths obtained from the $GW$ approximation relative to the exact solution obtained from Eq.~\ref{eq:gf}. Additionally, we see that the $GW$ quasiparticle peaks are in qualitatively better agreement with the exact solution, but that their relative peak strength is larger as a result of the underestimation of the satellite weights. 

\section{Conclusions and Outlook}

In summary, we have firmly demonstrated the connection between the electronic self-energy and coupled-cluster doubles theory. To do so, we decouple the particle-hole as well as forward- and backward-time sectors of the electronic self-energy. Our formal insights have demonstrated that CCD theory represents a self-consistent, infinite-order summation of fourth-order electronic self-energy diagrams restricted to the space of 2p1h/2h1p excitations. The relationship to the IP/EA-EOM-CCD eigenvalue problem can be revealed be coupling reverse-time self-energy contributions to either the occupied or virtual sectors while maintaining particle-hole separability. Finally, using the formalism developed in this paper, we reconstruct the exact Green's function of the Hubbard dimer in terms of the CCD amplitudes. We hope that our findings will help stimulate renewed work towards combining Green's function and coupled-cluster theories for ground and excited state many-body correlation. Further directions of this work would be to develop a hermitian ADC(3) self-energy approximation that contains the exact CCD amplitudes obtained from a ground state coupled-cluster calculation. This would allow for the use of ground state coupled-cluster amplitudes, that contain infinite-order partial summations of the forward-time self-energy diagrams derived in this work, within the hermitian structure of the conventional Green's function formalism to obtain charged excitation spectra. In principle, the relationships identified here can also be used to generate novel CCD approximations from the $GW$ approximation. The extension of the analysis presented here to the derivation of the coupled doubles and triples amplitude equations is reserved for future work. 

\begin{figure*}[ht]
    \centering
    \begin{gather*}
\begin{split}~\label{eq:ccd_se}
\bar{\Sigma}^{\text{CCD}}_{ij}(\omega) &= \hspace{2.5mm}
\begin{gathered}
    \begin{fmfgraph*}(40,60)
    \fmfcurved
    \fmfset{arrow_len}{3mm}
    \fmfleft{i1,i2}
    \fmflabel{}{i1}
    \fmflabel{}{i2}
    \fmfright{o1,o2}
    \fmflabel{}{o1}
    \fmflabel{}{o2}
    \fmf{fermion}{i1,i2}
    \fmf{wiggly}{o1,i1}
    \fmf{fermion,left=0.3,tension=0}{o1,o2}
    \fmf{wiggly}{o2,i2}
    \fmf{fermion,left=0.3,tension=0}{o2,o1}
    \fmfdot{o1,o2,i1,i2}
\end{fmfgraph*}
\end{gathered} \hspace{5mm}+\hspace{5mm} 
    \begin{gathered}
    \begin{fmfgraph*}(40,60)
    \fmfcurved
    \fmfset{arrow_len}{3mm}
    \fmfleft{i1,i2,i3}
    \fmflabel{}{i1}
    \fmflabel{}{i2}
    \fmfright{o1,o2,o3}
    \fmflabel{}{o1}
    \fmflabel{}{o2}
    \fmf{wiggly}{i2,o2}
    \fmf{wiggly}{i3,o3}
    \fmf{wiggly}{i1,o1}
    \fmf{fermion}{i1,i2}
    \fmf{fermion}{i2,i3}
    \fmf{fermion}{o1,o2}
    \fmf{fermion}{o2,o3}
    \fmf{fermion,left=0.3}{o3,o1}
    \fmfforce{(0.0w,0.0h)}{i1}
    \fmfforce{(1.0w,0.0h)}{o1}
    \fmfforce{(0.0w,1.0h)}{i3}
    \fmfforce{(1.0w,1.0h)}{o3}
    \fmfdot{i1,i2,i3}
    \fmfdot{o1,o2,o3}
\end{fmfgraph*}
\end{gathered}
\hspace{7.5mm}+\hspace{5mm}
\begin{gathered}
\begin{fmfgraph*}(40,60)
    \fmfcurved
    \fmfset{arrow_len}{3mm}
    \fmfleft{i1,i2,i3}
    \fmflabel{}{i1}
    \fmflabel{}{i2}
    \fmfright{o1,o2,o3}
    \fmflabel{}{o1}
    \fmflabel{}{o2}
    \fmf{wiggly}{i1,v1}
    \fmf{wiggly}{v1,o1}
    \fmf{wiggly}{v2,o2}
    \fmf{wiggly}{i3,v3}
    \fmf{fermion}{i1,i3}
    \fmf{fermion,left=0.3}{o1,o2}
    \fmf{fermion,left=0.3}{o2,o1}
    \fmf{fermion,left=0.3}{v2,v3}
    \fmf{fermion,left=0.3}{v3,v2}
    \fmfforce{(0.0w,0.0h)}{i1}
    \fmfforce{(1.0w,0.0h)}{o1}
    \fmfforce{(0.5w,0.5h)}{v2}
    \fmfforce{(0.5w,0.0h)}{v1}
    \fmfforce{(0.5w,1.0h)}{v3}
    \fmfforce{(0.0w,1.0h)}{i3}
    \fmfforce{(1.0w,1.0h)}{o3}
    \fmfdot{v2,v3}
    \fmfdot{i1,i3}
    \fmfdot{o1,o2}
\end{fmfgraph*}
\end{gathered}\hspace{7.5mm}+\hspace{5mm}
\begin{gathered}
\begin{fmfgraph*}(40,60)
    \fmfcurved
    \fmfset{arrow_len}{3mm}
    \fmfleft{i1,i2,i3}
    \fmflabel{}{i1}
    \fmflabel{}{i2}
    \fmfright{o1,o2,o3}
    \fmflabel{}{o1}
    \fmflabel{}{o2}
    \fmf{plain}{i1,o1}
    \fmf{wiggly}{i2,o2}
    \fmf{wiggly}{i3,o3}
    \fmf{fermion}{i2,i3}
    \fmf{fermion,left=0.3}{o3,i1}
    \fmf{fermion,left=0.3}{i1,o3}
    \fmf{fermion,left=0.3}{o2,o1}
    \fmf{fermion,left=0.3}{o1,o2}
    \fmfforce{(0.5w,0.0h)}{i1}
    \fmfforce{(1.0w,0.0h)}{o1}
    \fmfforce{(0.0w,0.5h)}{i2}
    \fmfforce{(1.0w,0.5h)}{o2}
    \fmfforce{(0.0w,1.0h)}{i3}
    \fmfforce{(0.5w,1.0h)}{o3}
    \fmfdot{i2,i3}
    \fmfdot{o2,o3}
\end{fmfgraph*}
\end{gathered}\hspace{5mm}+\hspace{5mm} 
    \begin{gathered}
    \begin{fmfgraph*}(40,60)
    \fmfcurved
    \fmfset{arrow_len}{3mm}
    \fmfleft{i1,i2,i3}
    \fmflabel{}{i1}
    \fmflabel{}{i2}
    \fmfright{o1,o2,o3}
    \fmflabel{}{o1}
    \fmflabel{}{o2}
    \fmf{plain}{i1,o1}
    \fmf{wiggly}{i2,o2}
    \fmf{wiggly}{i3,o3}
    \fmf{fermion}{i1,i3}
    \fmf{fermion}{i2,i1}
    \fmf{fermion,left=0.3}{o3,o2}
    \fmf{fermion}{o2,o1}
    \fmf{fermion}{o1,o3}
    \fmfforce{(0.0w,0.0h)}{i1}
    \fmfforce{(1.0w,0.0h)}{o1}
    \fmfforce{(0.0w,1.0h)}{i3}
    \fmfforce{(1.0w,1.0h)}{o3}
    \fmfforce{(0.75w,0.5h)}{i2}
    \fmfforce{(1.5w,0.5h)}{o2}
    \fmfdot{i2,i3}
    \fmfdot{o2,o3}
\end{fmfgraph*}
\end{gathered}\\
\\
&+\hspace{5mm}\begin{gathered}
    \begin{fmfgraph*}(60,80)
    \fmfcurved
    \fmfset{arrow_len}{3mm}
    \fmfleft{i1,i2,i3,i4}
    \fmflabel{}{i1}
    \fmflabel{}{i2}
    \fmfright{o1,o2,o3,o4}
    \fmflabel{}{o1}
    \fmflabel{}{o2}
    \fmf{wiggly}{i1,o1}
    \fmf{wiggly}{i3,o3}
    \fmf{plain}{i2,o2}
    \fmf{wiggly}{i4,o4}
    \fmf{fermion}{i1,i3}
    \fmf{fermion}{i3,i2}
    \fmf{fermion}{i2,i4}
    \fmf{fermion}{o1,o3}
    \fmf{fermion}{o3,o2}
    \fmf{fermion}{o2,o4}
    \fmf{fermion,left=0.3}{o4,o1}
    \fmfforce{(0.25w,0.0h)}{i1}
    \fmfforce{(1.0w,0.0h)}{o1}
    \fmfforce{(0.5w,0.66h)}{i3}
    \fmfforce{(1.1w,0.66h)}{o3}
    \fmfforce{(0.5w,0.66h)}{i3}
    \fmfforce{(0.0w,0.33h)}{i2}
    \fmfforce{(0.66w,0.33h)}{o2}
    \fmfforce{(0.0w,1.0h)}{i4}
    \fmfforce{(1.0w,1.0h)}{o4}
    \fmfdot{i1,i3,i4}
    \fmfdot{o1,o3,o4}
\end{fmfgraph*}
\end{gathered}
\hspace{7.5mm}+\hspace{5mm}
\begin{gathered}
\begin{fmfgraph*}(60,80)
    \fmfcurved
    \fmfset{arrow_len}{3mm}
    \fmfleft{i1,i2,i3,i4}
    \fmflabel{}{i1}
    \fmflabel{}{i2}
    \fmfright{o1,o2,o3,o4}
    \fmflabel{}{o1}
    \fmflabel{}{o2}
    \fmf{wiggly}{i1,o1}
    \fmf{plain}{i2,o2}
    \fmf{wiggly}{i4,o4}
    \fmf{wiggly}{i3,o3}
    \fmf{fermion}{i1,i4}
    \fmf{fermion,left=0.3}{o1,o3}
    \fmf{fermion,left=0.3}{o3,o1}
    \fmf{fermion,left=0.3}{o2,o4}
    \fmf{fermion,left=0.3}{o4,o2}
    \fmf{fermion,left=0.3}{i2,i3}
    \fmf{fermion,left=0.3}{i3,i2}
    \fmfforce{(0.0w,0.0h)}{i1}
    \fmfforce{(1.0w,0.0h)}{o1}
    \fmfforce{(0.0w,1.0h)}{i4}
    \fmfforce{(0.66w,1.0h)}{o4}
    \fmfforce{(0.66w,0.33h)}{o2}
    \fmfforce{(0.33w,0.33h)}{i2}
    \fmfforce{(0.33w,0.66h)}{i3}
    \fmfforce{(1.0w,0.66h)}{o3}
    \fmfdot{i1,i3,i4}
    \fmfdot{o1,o3,o4}
\end{fmfgraph*}
\end{gathered}\hspace{5mm}+\hspace{10mm}
\begin{gathered}
    \begin{fmfgraph*}(40,60)
    \fmfcurved
    \fmfset{arrow_len}{3mm}
    \fmfleft{i1,i2}
    \fmflabel{}{i1}
    \fmflabel{}{i2}
    \fmfright{o1,o2}
    \fmflabel{}{o1}
    \fmflabel{}{o2}
    \fmf{fermion}{i1,v4}
    \fmf{fermion}{v4,v3}
    \fmf{fermion}{v3,i2}
    \fmf{wiggly}{o1,i1}
    \fmf{plain}{v4,v1}
    \fmf{wiggly}{v3,v2}
    \fmf{fermion,left=0.3}{v1,v2}
    \fmf{fermion,left=0.3}{v2,v1}
    \fmf{fermion,left=0.3,tension=0}{o1,o2}
    \fmf{wiggly}{o2,i2}
    \fmf{fermion,left=0.3,tension=0}{o2,o1}
    \fmfdot{o1,o2,i1,i2}
    \fmfdot{v3,v2}
    \fmfforce{(0.2w,0.5h)}{i1}
    \fmfforce{(1.0w,0.5h)}{o1}
    \fmfforce{(0.0w,1.0h)}{i2}
    \fmfforce{(1.0w,1.0h)}{o2}
    \fmfforce{(0.0w,0.33h)}{v4}
    \fmfforce{(0.0w,0.66h)}{v3}
    \fmfforce{(-0.33w,0.33h)}{v1}
    \fmfforce{(-0.33w,0.66h)}{v2}
\end{fmfgraph*}
\end{gathered}\hspace{5mm}+\hspace{5mm}
\begin{gathered}
    \begin{fmfgraph*}(40,60)
    \fmfcurved
    \fmfset{arrow_len}{3mm}
    \fmfleft{i1,i2}
    \fmflabel{}{i1}
    \fmflabel{}{i2}
    \fmfright{o1,o2}
    \fmflabel{}{o1}
    \fmflabel{}{o2}
    \fmf{fermion}{i1,i2}
    \fmf{fermion,left=0.3}{o2,v3}
    \fmf{fermion,left=0.3}{v3,v4}
    \fmf{fermion}{v4,o1}
    \fmf{wiggly}{o1,i1}
    \fmf{plain}{v4,v1}
    \fmf{wiggly}{v3,v2}
    \fmf{fermion,left=0.3}{v1,v2}
    \fmf{fermion,left=0.3}{v2,v1}
    \fmf{fermion}{o1,o2}
    \fmf{wiggly}{o2,i2}
    \fmfdot{o1,o2,i1,i2}
    \fmfdot{v3,v2}
    \fmfforce{(0.0w,0.5h)}{i1}
    \fmfforce{(1.0w,0.5h)}{o1}
    \fmfforce{(1.0w,1.0h)}{o2}
    \fmfforce{(0.0w,1.0h)}{i2}
    \fmfforce{(1.33w,0.33h)}{v4}
    \fmfforce{(1.33w,0.66h)}{v3}
    \fmfforce{(1.8w,0.33h)}{v1}
    \fmfforce{(1.8w,0.66h)}{v2}
\end{fmfgraph*}
\end{gathered}\hspace{15mm}+\hspace{2.5mm}\cdots
\end{split}
\end{gather*}
\caption{The 2p1h/2h1p excitation restricted Feynman-Goldstone one-particle irreducible electronic particle-hole-time decoupled self-energy diagrams that generate the CCD amplitude equations when summed to infinite-order by the Dyson supermatrix. The final four diagrams depicted are fourth-order self-energy contributions. The solid line represents the CCD amplitude, $t^{ab}_{ij}$.}
    \label{fig:ccd_se}
\end{figure*}
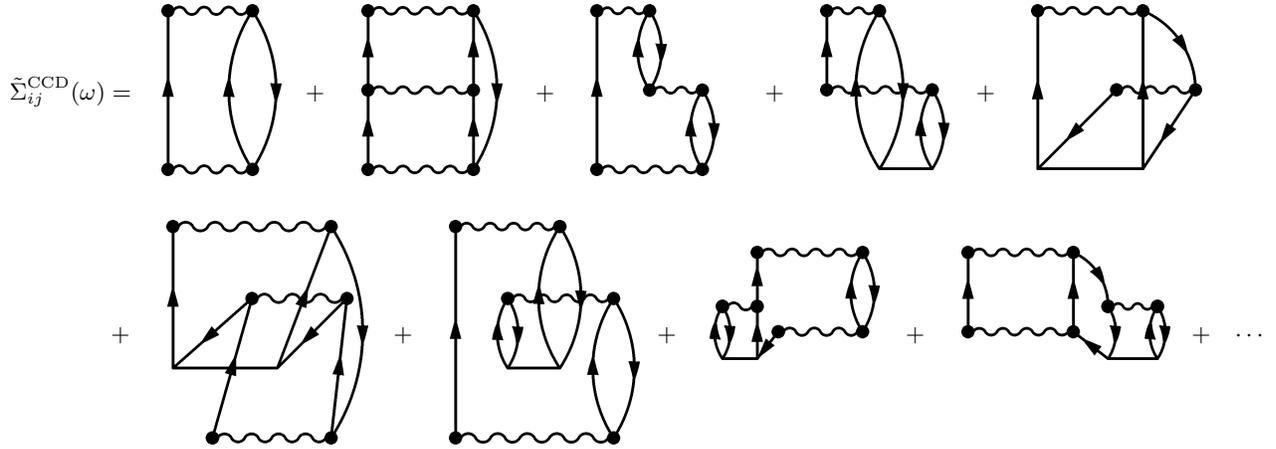
\section*{Acknowledgments}

The author is grateful to D. P. Tew for supporting this work and providing insightful comments on the manuscript. 

\appendix

\section{Particle-hole-time decoupled electronic self-energy diagrams contained in CCD theory}~\label{app:2}

The set of one-particle irreducible particle-hole-time decoupled self-energy diagrams that generate the exact CCD amplitude equations are shown in Fig.~\ref{fig:ccd_se}. In Fig.~\ref{fig:ccd_se}, we introduce the self-consistent notation, whereby the interaction and coupling matrices are now dependent on the solution of the CCD amplitudes. This amounts to replacing the MP2 amplitudes that are be obtained from the perturbative electronic self-energy diagrams by the full CCD amplitudes: $(t^{ab}_{ij})_{\text{MP2}}\rightarrow t^{ab}_{ij}$. 

Downfolding Eq.~\ref{eq:ccd_super}, using the modified self-consistent coupling and interaction matrices that depend on the exact CCD amplitudes (Eqs~\ref{eq:sc_ints} and~\ref{eq:mod_ints}) leads to the expression
\begin{gather}
    \begin{split}~\label{eq:se_ccd}
        &\bar{\Sigma}^{\text{CCD}}_{ij}(\omega) = \\
        &\frac{1}{4}\sum_{\substack{kab\\lcd}}\bar{U}_{i,abk}\Big((\omega+i\eta)\mathbbm{1}-(\mathbf{\bar{K}}^{>,2\p1\h}+\mathbf{\bar{C}}^{>,2\p1\h})\Big)^{-1}_{kab,lcd} U^{\text{sc}}_{cdl,j} \ ,
    \end{split}
\end{gather}
where the interaction matrix elements, $(\mathbf{\bar{K}}^{>,2\p1\h}_{kab,jcd}+\mathbf{\bar{C}}^{>,2\p1\h}_{kab,jcd})$, are obtained by using Eqs~\ref{eq:mod_ints} to update the interaction matrices defined in Eq.~\ref{eq:forward}. The self-consistent coupling matrix elements, $U^{\text{sc}}_{cdl,j}$, are defined in Eq.~\ref{eq:sc_ints}. The infinite series of self-energy diagrams contained in Eq.~\ref{eq:se_ccd} is presented in Fig.~\ref{fig:ccd_se}.

\section{Doubles amplitudes from the particle-hole-time decoupled self-energy of the virtual states}~\label{app:1}

For the virtual block of the particle-hole-time decoupled self-energy, we have the following supermatrix eigenvalue problem 
\begin{gather}
    \begin{split}
         \left(\begin{array}{cc}
            \mathbf{f} & \mathbf{\bar{V}} \\
            \mathbf{V}^\dag & \mathbf{K}^{<,2\h1\p}+\mathbf{C}^{<,2\h1\p} 
        \end{array}\right)\left(\begin{array}{c}
             \mathbf{\tilde{X}} \\
              \mathbf{\tilde{Y}}
        \end{array}\right) = \left(\begin{array}{c}
             \mathbf{\tilde{X}} \\
              \mathbf{\tilde{Y}}
        \end{array}\right)\mathbf{\tilde{\mathcal{E}}} \ .
    \end{split}
\end{gather}
Defining the effective doubles amplitudes as $\mathbf{\tilde{T}}=\mathbf{\tilde{Y}}\mathbf{\tilde{X}}^{-1}$ we have the extended Fock operator and self-energy Riccati equation 
\begin{subequations}
    \begin{align}
    \begin{split}
        \mathbf{F}\mathbf{\tilde{X}} &= \mathbf{\tilde{X}}\mathbf{\tilde{\mathcal{E}}}
    \end{split}\\
    \begin{split}~\label{eq:Riccati}
    \mathbf{V}^\dag + (\mathbf{K}^{<,2\h1\p}+\mathbf{C}^{<,2\h1\p})\mathbf{\tilde{T}} -\mathbf{\tilde{T}}\mathbf{F} &= \mathbf{0} \ ,
    \end{split}
\end{align}
\end{subequations}
where $\mathbf{F}=\mathbf{f} + \mathbf{\bar{V}}\mathbf{\tilde{T}}$. We immediately apply the constraint of particle-hole symmetry: $\Tilde{t}^{ab}_{ij}= -(t^{ab}_{ij})^*$. The corresponding extended Fock operator in the virtual space is then given by 
\begin{gather}
    \begin{split}~\label{eq:vir_fock}
        F_{ab} &= \epsilon_{a}\delta_{ab} + \frac{1}{2}\sum_{klc}\braket{ac||kl}\Tilde{t}^{bc}_{kl}\\
        &= \epsilon_{a}\delta_{ab} - \frac{1}{2}\sum_{klc}\braket{kl||bc}t^{ac}_{kl} \ ,
    \end{split}
\end{gather}
where we have used the constraint of hermiticity: $F_{ab} = F^*_{ba}$.
This expression is exactly that obtained from CCD theory for the extended Fock operator~\cite{coveney2023coupled,coveney2024cc_se}. The amplitude equations become 
\begin{gather}
    \begin{split}
        &\braket{ij||ab} +\frac{1}{2}\sum_{cd} \braket{ab||cd}(t^{cd}_{ij})_{\text{MP$2$}} +\sum_{kc}\braket{ka||jc}(t^{bc}_{ik})_{\text{MP$2$}}\\
        &-\sum_{kc}\braket{ak||ci}(t^{bc}_{jk})_{\text{MP$2$}}
        -\Delta^{ab}_{ij}\tilde{t}^{ab}_{ij} - \frac{1}{2}\sum_{kl}\braket{ij||kl}\tilde{t}^{ab}_{kl} \\
        &+\sum_{kc}\braket{jc||kb}\tilde{t}^{ac}_{ik}-\sum_{kc}\braket{ic||kb}\tilde{t}^{ac}_{jk}\\
        &-\frac{1}{2}\sum_{klcd}\tilde{t}^{ac}_{kl          }\braket{cd||kl}\tilde{t}^{bd}_{ij} = 0 \ ,
    \end{split}
\end{gather}
which can be re-written by imposing the particle-hole symmetry constraint as
\begin{gather}
    \begin{split}
        &\braket{ab||ij} +\frac{1}{2}\sum_{cd} \braket{ab||cd}(t^{cd}_{ij})_{\text{MP$2$}} +\sum_{kc}\braket{ka||jc}(t^{bc}_{ik})_{\text{MP$2$}}\\
        &-\sum_{kc}\braket{ak||ci}(t^{bc}_{jk})_{\text{MP$2$}} +\Delta^{ab}_{ij}t^{ab}_{ij} +\frac{1}{2}\sum_{kl}\braket{ij||kl}t^{ab}_{kl} \\
        &-\sum_{kc}\braket{kb||jc}t^{ac}_{ik}+\sum_{kc}\braket{kb||ic}t^{ac}_{jk}\\
        &-\frac{1}{2}\sum_{klcd}t^{ac}_{kl}\braket{kl||cd}t^{bd}_{ij} = 0 \ .
    \end{split}
\end{gather}
To generate the full CCD equations from the virtual block, we must self-consistently update the coupling and interaction matrices as follows. First, we modify the coupling matrices to be self-consistently determined from the doubles amplitude solutions:
\begin{gather}
    \begin{split}
        V^{\text{sc}}_{p,kla} &=  \braket{pa||kl} + \sum_{cm}\braket{pm||kc}t^{ca}_{ml} -\sum_{cm}\braket{pm||lc}t^{ca}_{mk} \\
        &+\frac{1}{2}\sum_{cd} \braket{pa||cd}t^{cd}_{kl} \ .
        \end{split}
\end{gather}
This is analogous to the procedure carried out for the occupied block. Then we update the interaction matrices between the 2h1p states to become self-consistently dependent on the doubles amplitudes as
\begin{subequations}
    \begin{align}
    \begin{split}
        \epsilon_a\delta_{ab} &\rightarrow \epsilon_a\delta_{ab} -\frac{1}{2}\sum_{ijc} \braket{ij||bc}t^{ca}_{ij} = F_{ab}
    \end{split}\\
    \begin{split}
        \epsilon_i\delta_{ij} &\rightarrow \epsilon_i\delta_{ij} +\frac{1}{2}\sum_{kcd} \braket{ik||cd}t^{cd}_{jk} = F_{ij}
        \end{split}\\
    \begin{split}
        \braket{ij||kl} &\rightarrow \braket{ij||kl} +\frac{1}{2}\sum_{cd}\braket{ij||cd} t^{cd}_{kl} = \chi_{ij,kl}
         \end{split}\\
    \begin{split}
        \braket{ja||bi} &\rightarrow \braket{ja||bi} + \sum_{kc}  \braket{jk||bc}t^{ac}_{ik} = \chi_{ja,bi}\ .
    \end{split}
\end{align}
\end{subequations}
Using these quantities, we write the modified self-consistent interaction matrices as 
\begin{gather}
    \begin{split}
        (\mathbf{\bar{K}}^{<,2\h1\p}_{ija,klb}+\mathbf{\bar{C}}^{<,2\h1\p}_{ija,klb}) &= F_{ik}\delta_{ab}\delta_{jl}+ F_{jl}\delta_{ab}\delta_{ik}-F_{ba}\delta_{ik}\delta_{jl} \\
        &- F_{il}\delta_{ab}\delta_{jk}- F_{jk}\delta_{ab}\delta_{il}+F_{ba}\delta_{il}\delta_{jk}\\&
        +\chi_{ib,al}\delta_{jk}+\chi_{jb,ak}\delta_{il}-\chi_{ij,kl}\delta_{ab} \\
    &-\chi_{ib,ak}\delta_{jl}- \chi_{jb,al}\delta_{ik} \ .
    \end{split}
\end{gather}
Resolving these equations in the self-energy Riccati equation (Eq.~\ref{eq:Riccati}) yields the full CCD amplitude equations.

\end{fmffile}

\section*{References}
\bibliography{qpmp}

\begin{thebibliography}{56}%
\makeatletter
\providecommand \@ifxundefined [1]{%
 \@ifx{#1\undefined}
}%
\providecommand \@ifnum [1]{%
 \ifnum #1\expandafter \@firstoftwo
 \else \expandafter \@secondoftwo
 \fi
}%
\providecommand \@ifx [1]{%
 \ifx #1\expandafter \@firstoftwo
 \else \expandafter \@secondoftwo
 \fi
}%
\providecommand \natexlab [1]{#1}%
\providecommand \enquote  [1]{``#1''}%
\providecommand \bibnamefont  [1]{#1}%
\providecommand \bibfnamefont [1]{#1}%
\providecommand \citenamefont [1]{#1}%
\providecommand \href@noop [0]{\@secondoftwo}%
\providecommand \href [0]{\begingroup \@sanitize@url \@href}%
\providecommand \@href[1]{\@@startlink{#1}\@@href}%
\providecommand \@@href[1]{\endgroup#1\@@endlink}%
\providecommand \@sanitize@url [0]{\catcode `\\12\catcode `\$12\catcode
  `\&12\catcode `\#12\catcode `\^12\catcode `\_12\catcode `\%12\relax}%
\providecommand \@@startlink[1]{}%
\providecommand \@@endlink[0]{}%
\providecommand \url  [0]{\begingroup\@sanitize@url \@url }%
\providecommand \@url [1]{\endgroup\@href {#1}{\urlprefix }}%
\providecommand \urlprefix  [0]{URL }%
\providecommand \Eprint [0]{\href }%
\providecommand \doibase [0]{http://dx.doi.org/}%
\providecommand \selectlanguage [0]{\@gobble}%
\providecommand \bibinfo  [0]{\@secondoftwo}%
\providecommand \bibfield  [0]{\@secondoftwo}%
\providecommand \translation [1]{[#1]}%
\providecommand \BibitemOpen [0]{}%
\providecommand \bibitemStop [0]{}%
\providecommand \bibitemNoStop [0]{.\EOS\space}%
\providecommand \EOS [0]{\spacefactor3000\relax}%
\providecommand \BibitemShut  [1]{\csname bibitem#1\endcsname}%
\let\auto@bib@innerbib\@empty
\bibitem [{\citenamefont {Fetter}\ and\ \citenamefont
  {Walecka}(1971)}]{Quantum}%
  \BibitemOpen
  \bibfield  {author} {\bibinfo {author} {\bibfnamefont {A}~\bibnamefont
  {Fetter}}\ and\ \bibinfo {author} {\bibfnamefont {J~D}\ \bibnamefont
  {Walecka}},\ }\href@noop {} {\emph {\bibinfo {title} {Quantum theory of many
  particle systems}}}\ (\bibinfo  {publisher} {McGraw-Hill},\ \bibinfo {year}
  {1971})\BibitemShut {NoStop}%
\bibitem [{\citenamefont {Stefanucci}\ and\ \citenamefont
  {Van~Leeuwen}(2013)}]{stefanucci2013nonequilibrium}%
  \BibitemOpen
  \bibfield  {author} {\bibinfo {author} {\bibfnamefont {G}~\bibnamefont
  {Stefanucci}}\ and\ \bibinfo {author} {\bibfnamefont {R}~\bibnamefont
  {Van~Leeuwen}},\ }\href@noop {} {\emph {\bibinfo {title} {Nonequilibrium
  many-body theory of quantum systems: a modern introduction}}}\ (\bibinfo
  {publisher} {Cambridge University Press},\ \bibinfo {year}
  {2013})\BibitemShut {NoStop}%
\bibitem [{\citenamefont {Mahan}(2000)}]{mahan2000many}%
  \BibitemOpen
  \bibfield  {author} {\bibinfo {author} {\bibfnamefont {G~D}\ \bibnamefont
  {Mahan}},\ }\href@noop {} {\emph {\bibinfo {title} {Many-particle physics}}}\
  (\bibinfo  {publisher} {Springer Science \& Business Media},\ \bibinfo {year}
  {2000})\BibitemShut {NoStop}%
\bibitem [{\citenamefont {Quintero-Monsebaiz}\ \emph
  {et~al.}(2022)\citenamefont {Quintero-Monsebaiz}, \citenamefont {Monino},
  \citenamefont {Marie},\ and\ \citenamefont {Loos}}]{quintero2022connections}%
  \BibitemOpen
  \bibfield  {author} {\bibinfo {author} {\bibfnamefont {R}~\bibnamefont
  {Quintero-Monsebaiz}}, \bibinfo {author} {\bibfnamefont {E}~\bibnamefont
  {Monino}}, \bibinfo {author} {\bibfnamefont {A}~\bibnamefont {Marie}}, \ and\
  \bibinfo {author} {\bibfnamefont {P-F}\ \bibnamefont {Loos}},\ }\bibfield
  {title} {\enquote {\bibinfo {title} {Connections between many-body
  perturbation and coupled-cluster theories},}\ }\href
  {https://doi.org/10.1063/5.0130837} {\bibfield  {journal} {\bibinfo
  {journal} {The Journal of Chemical Physics}\ }\textbf {\bibinfo {volume}
  {157}},\ \bibinfo {pages} {231102} (\bibinfo {year} {2022})}\BibitemShut
  {NoStop}%
\bibitem [{\citenamefont {Monino}\ and\ \citenamefont
  {Loos}(2023)}]{monino2023connections}%
  \BibitemOpen
  \bibfield  {author} {\bibinfo {author} {\bibfnamefont {E}~\bibnamefont
  {Monino}}\ and\ \bibinfo {author} {\bibfnamefont {P-F}\ \bibnamefont
  {Loos}},\ }\bibfield  {title} {\enquote {\bibinfo {title} {Connections and
  performances of {Green’s} function methods for charged and neutral
  excitations},}\ }\href {https://doi.org/10.1063/5.0159853} {\bibfield
  {journal} {\bibinfo  {journal} {The Journal of Chemical Physics}\ }\textbf
  {\bibinfo {volume} {159}},\ \bibinfo {pages} {034105} (\bibinfo {year}
  {2023})}\BibitemShut {NoStop}%
\bibitem [{\citenamefont {Berkelbach}(2018)}]{berkelbach2018communication}%
  \BibitemOpen
  \bibfield  {author} {\bibinfo {author} {\bibfnamefont {T~C}\ \bibnamefont
  {Berkelbach}},\ }\bibfield  {title} {\enquote {\bibinfo {title}
  {Communication: {Random-Phase} approximation excitation energies from
  approximate equation-of-motion coupled-cluster doubles},}\ }\href
  {https://doi.org/10.1063/1.5032314} {\bibfield  {journal} {\bibinfo
  {journal} {The Journal of Chemical Physics}\ }\textbf {\bibinfo {volume}
  {149}},\ \bibinfo {pages} {041103} (\bibinfo {year} {2018})}\BibitemShut
  {NoStop}%
\bibitem [{\citenamefont {T{\"o}lle}\ and\ \citenamefont
  {Chan}(2023)}]{tolle2023exact}%
  \BibitemOpen
  \bibfield  {author} {\bibinfo {author} {\bibfnamefont {J}~\bibnamefont
  {T{\"o}lle}}\ and\ \bibinfo {author} {\bibfnamefont {G~K-L}\ \bibnamefont
  {Chan}},\ }\bibfield  {title} {\enquote {\bibinfo {title} {Exact
  relationships between the {$GW$} approximation and equation-of-motion
  coupled-cluster theories through the quasi-boson formalism},}\ }\href
  {https://doi.org/10.1063/5.0139716} {\bibfield  {journal} {\bibinfo
  {journal} {The Journal of Chemical Physics}\ }\textbf {\bibinfo {volume}
  {158}},\ \bibinfo {pages} {124123} (\bibinfo {year} {2023})}\BibitemShut
  {NoStop}%
\bibitem [{\citenamefont {Bintrim}\ and\ \citenamefont
  {Berkelbach}(2021)}]{bintrim2021full}%
  \BibitemOpen
  \bibfield  {author} {\bibinfo {author} {\bibfnamefont {S~J}\ \bibnamefont
  {Bintrim}}\ and\ \bibinfo {author} {\bibfnamefont {T~C}\ \bibnamefont
  {Berkelbach}},\ }\bibfield  {title} {\enquote {\bibinfo {title}
  {Full-frequency {$GW$} without frequency},}\ }\href
  {https://doi.org/10.1063/5.0035141} {\bibfield  {journal} {\bibinfo
  {journal} {The Journal of Chemical Physics}\ }\textbf {\bibinfo {volume}
  {154}},\ \bibinfo {pages} {041101} (\bibinfo {year} {2021})}\BibitemShut
  {NoStop}%
\bibitem [{\citenamefont {Bintrim}\ and\ \citenamefont
  {Berkelbach}(2022)}]{bintrim2022full}%
  \BibitemOpen
  \bibfield  {author} {\bibinfo {author} {\bibfnamefont {S~J}\ \bibnamefont
  {Bintrim}}\ and\ \bibinfo {author} {\bibfnamefont {T~C}\ \bibnamefont
  {Berkelbach}},\ }\bibfield  {title} {\enquote {\bibinfo {title}
  {Full-frequency dynamical {Bethe--Salpeter} equation without frequency and a
  study of double excitations},}\ }\href {https://doi.org/10.1063/5.0074434}
  {\bibfield  {journal} {\bibinfo  {journal} {The Journal of Chemical Physics}\
  }\textbf {\bibinfo {volume} {156}},\ \bibinfo {pages} {044114} (\bibinfo
  {year} {2022})}\BibitemShut {NoStop}%
\bibitem [{\citenamefont {Lange}\ and\ \citenamefont
  {Berkelbach}(2018)}]{lange2018relation}%
  \BibitemOpen
  \bibfield  {author} {\bibinfo {author} {\bibfnamefont {M~F}\ \bibnamefont
  {Lange}}\ and\ \bibinfo {author} {\bibfnamefont {T~C}\ \bibnamefont
  {Berkelbach}},\ }\bibfield  {title} {\enquote {\bibinfo {title} {On the
  relation between equation-of-motion coupled-cluster theory and the {$GW$}
  approximation},}\ }\href {https://doi.org/10.1021/acs.jctc.8b00455}
  {\bibfield  {journal} {\bibinfo  {journal} {Journal of Chemical Theory and
  Computation}\ }\textbf {\bibinfo {volume} {14}},\ \bibinfo {pages}
  {4224--4236} (\bibinfo {year} {2018})}\BibitemShut {NoStop}%
\bibitem [{\citenamefont {Opoku}\ \emph {et~al.}(2021)\citenamefont {Opoku},
  \citenamefont {Paw{\l}owski},\ and\ \citenamefont {Ortiz}}]{opoku2021new}%
  \BibitemOpen
  \bibfield  {author} {\bibinfo {author} {\bibfnamefont {E}~\bibnamefont
  {Opoku}}, \bibinfo {author} {\bibfnamefont {F}~\bibnamefont {Paw{\l}owski}},
  \ and\ \bibinfo {author} {\bibfnamefont {J~V}\ \bibnamefont {Ortiz}},\
  }\bibfield  {title} {\enquote {\bibinfo {title} {A new generation of diagonal
  self-energies for the calculation of electron removal energies},}\ }\href
  {https://doi.org/10.1063/5.0070849} {\bibfield  {journal} {\bibinfo
  {journal} {The Journal of Chemical Physics}\ }\textbf {\bibinfo {volume}
  {155}},\ \bibinfo {pages} {204107} (\bibinfo {year} {2021})}\BibitemShut
  {NoStop}%
\bibitem [{\citenamefont {Opoku}\ \emph
  {et~al.}(2023{\natexlab{a}})\citenamefont {Opoku}, \citenamefont
  {Paw{\l}owski},\ and\ \citenamefont {Ortiz}}]{opoku2023new}%
  \BibitemOpen
  \bibfield  {author} {\bibinfo {author} {\bibfnamefont {E}~\bibnamefont
  {Opoku}}, \bibinfo {author} {\bibfnamefont {F}~\bibnamefont {Paw{\l}owski}},
  \ and\ \bibinfo {author} {\bibfnamefont {J~V}\ \bibnamefont {Ortiz}},\
  }\bibfield  {title} {\enquote {\bibinfo {title} {A new generation of
  non-diagonal, renormalized self-energies for calculation of electron removal
  energies},}\ }\href {https://doi.org/10.1063/5.0168779} {\bibfield  {journal}
  {\bibinfo  {journal} {The Journal of Chemical Physics}\ }\textbf {\bibinfo
  {volume} {159}},\ \bibinfo {pages} {124109} (\bibinfo {year}
  {2023}{\natexlab{a}})}\BibitemShut {NoStop}%
\bibitem [{\citenamefont {Opoku}\ \emph
  {et~al.}(2023{\natexlab{b}})\citenamefont {Opoku}, \citenamefont
  {Paw{\l}owski},\ and\ \citenamefont {Ortiz}}]{opoku2023new_prop}%
  \BibitemOpen
  \bibfield  {author} {\bibinfo {author} {\bibfnamefont {E}~\bibnamefont
  {Opoku}}, \bibinfo {author} {\bibfnamefont {F}~\bibnamefont {Paw{\l}owski}},
  \ and\ \bibinfo {author} {\bibfnamefont {J~V}\ \bibnamefont {Ortiz}},\
  }\bibfield  {title} {\enquote {\bibinfo {title} {New-generation
  electron-propagator methods for calculations of electron affinities and
  ionization energies: Tests on organic photovoltaic molecules},}\ }\href
  {https://doi.org/10.1021/acs.jctc.3c00954} {\bibfield  {journal} {\bibinfo
  {journal} {Journal of Chemical Theory and Computation}\ }\textbf {\bibinfo
  {volume} {20}},\ \bibinfo {pages} {290--306} (\bibinfo {year}
  {2023}{\natexlab{b}})}\BibitemShut {NoStop}%
\bibitem [{\citenamefont {Opoku}\ \emph {et~al.}(2024)\citenamefont {Opoku},
  \citenamefont {Paw{\l}owski},\ and\ \citenamefont
  {Ortiz}}]{opoku2024new_prop}%
  \BibitemOpen
  \bibfield  {author} {\bibinfo {author} {\bibfnamefont {E}~\bibnamefont
  {Opoku}}, \bibinfo {author} {\bibfnamefont {F}~\bibnamefont {Paw{\l}owski}},
  \ and\ \bibinfo {author} {\bibfnamefont {J~V}\ \bibnamefont {Ortiz}},\
  }\bibfield  {title} {\enquote {\bibinfo {title} {New-generation
  electron-propagator methods for molecular electron-binding energies},}\
  }\href {https://doi.org/10.1021/acs.jpca.3c08455} {\bibfield  {journal}
  {\bibinfo  {journal} {The Journal of Physical Chemistry A}\ }\textbf
  {\bibinfo {volume} {128}},\ \bibinfo {pages} {1399--1416} (\bibinfo {year}
  {2024})}\BibitemShut {NoStop}%
\bibitem [{\citenamefont {Coveney}\ and\ \citenamefont
  {Tew}(2025{\natexlab{a}})}]{coveney2023coupled}%
  \BibitemOpen
  \bibfield  {author} {\bibinfo {author} {\bibfnamefont {C~J~N}\ \bibnamefont
  {Coveney}}\ and\ \bibinfo {author} {\bibfnamefont {D~P}\ \bibnamefont
  {Tew}},\ }\bibfield  {title} {\enquote {\bibinfo {title} {Diagrammatic theory
  of the irreducible coupled-cluster self-energy},}\ }\href
  {https://doi.org/10.1103/p41w-bl6p} {\bibfield  {journal} {\bibinfo
  {journal} {Physical Review B}\ }\textbf {\bibinfo {volume} {112}},\ \bibinfo
  {pages} {045104} (\bibinfo {year} {2025}{\natexlab{a}})}\BibitemShut
  {NoStop}%
\bibitem [{\citenamefont {Coveney}\ and\ \citenamefont
  {Tew}(2025{\natexlab{b}})}]{coveney2024cc_se}%
  \BibitemOpen
  \bibfield  {author} {\bibinfo {author} {\bibfnamefont {C~J~N}\ \bibnamefont
  {Coveney}}\ and\ \bibinfo {author} {\bibfnamefont {D~P}\ \bibnamefont
  {Tew}},\ }\bibfield  {title} {\enquote {\bibinfo {title} {Non-hermitian
  {Green's} function theory with {$N$-body} interactions: the coupled-cluster
  similarity transformation},}\ }\href {https://arxiv.org/abs/2503.06586}
  {\bibfield  {journal} {\bibinfo  {journal} {arXiv preprint arXiv:2503.06586}\
  } (\bibinfo {year} {2025}{\natexlab{b}})}\BibitemShut {NoStop}%
\bibitem [{\citenamefont {Schirmer}\ \emph {et~al.}(1983)\citenamefont
  {Schirmer}, \citenamefont {Cederbaum},\ and\ \citenamefont
  {Walter}}]{schirmer1983new}%
  \BibitemOpen
  \bibfield  {author} {\bibinfo {author} {\bibfnamefont {J}~\bibnamefont
  {Schirmer}}, \bibinfo {author} {\bibfnamefont {L~S}\ \bibnamefont
  {Cederbaum}}, \ and\ \bibinfo {author} {\bibfnamefont {O}~\bibnamefont
  {Walter}},\ }\bibfield  {title} {\enquote {\bibinfo {title} {New approach to
  the one-particle green's function for finite fermi systems},}\ }\href
  {https://doi.org/10.1103/PhysRevA.28.1237} {\bibfield  {journal} {\bibinfo
  {journal} {Physical Review A}\ }\textbf {\bibinfo {volume} {28}},\ \bibinfo
  {pages} {1237} (\bibinfo {year} {1983})}\BibitemShut {NoStop}%
\bibitem [{\citenamefont {Caruso}\ \emph {et~al.}(2013)\citenamefont {Caruso},
  \citenamefont {Rinke}, \citenamefont {Ren}, \citenamefont {Rubio},\ and\
  \citenamefont {Scheffler}}]{caruso2013self}%
  \BibitemOpen
  \bibfield  {author} {\bibinfo {author} {\bibfnamefont {F}~\bibnamefont
  {Caruso}}, \bibinfo {author} {\bibfnamefont {P}~\bibnamefont {Rinke}},
  \bibinfo {author} {\bibfnamefont {X}~\bibnamefont {Ren}}, \bibinfo {author}
  {\bibfnamefont {A}~\bibnamefont {Rubio}}, \ and\ \bibinfo {author}
  {\bibfnamefont {M}~\bibnamefont {Scheffler}},\ }\bibfield  {title} {\enquote
  {\bibinfo {title} {Self-consistent {$GW$}: All-electron implementation with
  localized basis functions},}\ }\href
  {https://doi.org/10.1103/PhysRevB.88.075105} {\bibfield  {journal} {\bibinfo
  {journal} {Physical Review B}\ }\textbf {\bibinfo {volume} {88}},\ \bibinfo
  {pages} {075105} (\bibinfo {year} {2013})}\BibitemShut {NoStop}%
\bibitem [{\citenamefont {Raimondi}\ and\ \citenamefont
  {Barbieri}(2018)}]{raimondi2018algebraic}%
  \BibitemOpen
  \bibfield  {author} {\bibinfo {author} {\bibfnamefont {F}~\bibnamefont
  {Raimondi}}\ and\ \bibinfo {author} {\bibfnamefont {C}~\bibnamefont
  {Barbieri}},\ }\bibfield  {title} {\enquote {\bibinfo {title} {Algebraic
  diagrammatic construction formalism with three-body interactions},}\ }\href
  {https://doi.org/10.1103/PhysRevC.97.054308} {\bibfield  {journal} {\bibinfo
  {journal} {Physical Review C}\ }\textbf {\bibinfo {volume} {97}},\ \bibinfo
  {pages} {054308} (\bibinfo {year} {2018})}\BibitemShut {NoStop}%
\bibitem [{\citenamefont {Schirmer}(2018)}]{schirmer2018many}%
  \BibitemOpen
  \bibfield  {author} {\bibinfo {author} {\bibfnamefont {J}~\bibnamefont
  {Schirmer}},\ }\href@noop {} {\emph {\bibinfo {title} {Many-body methods for
  atoms, molecules and clusters}}},\ Vol.~\bibinfo {volume} {94}\ (\bibinfo
  {publisher} {Springer},\ \bibinfo {year} {2018})\BibitemShut {NoStop}%
\bibitem [{\citenamefont {Backhouse}\ \emph {et~al.}(2020)\citenamefont
  {Backhouse}, \citenamefont {Nusspickel},\ and\ \citenamefont
  {Booth}}]{backhouse2020wave}%
  \BibitemOpen
  \bibfield  {author} {\bibinfo {author} {\bibfnamefont {O~J}\ \bibnamefont
  {Backhouse}}, \bibinfo {author} {\bibfnamefont {M}~\bibnamefont
  {Nusspickel}}, \ and\ \bibinfo {author} {\bibfnamefont {G~H}\ \bibnamefont
  {Booth}},\ }\bibfield  {title} {\enquote {\bibinfo {title} {Wave function
  perspective and efficient truncation of renormalized second-order
  perturbation theory},}\ }\href {https://doi.org/10.1021/acs.jctc.9b01182}
  {\bibfield  {journal} {\bibinfo  {journal} {Journal of Chemical Theory and
  Computation}\ }\textbf {\bibinfo {volume} {16}},\ \bibinfo {pages}
  {1090--1104} (\bibinfo {year} {2020})}\BibitemShut {NoStop}%
\bibitem [{\citenamefont {Scott}\ \emph {et~al.}(2023)\citenamefont {Scott},
  \citenamefont {Backhouse},\ and\ \citenamefont {Booth}}]{scott2023moment}%
  \BibitemOpen
  \bibfield  {author} {\bibinfo {author} {\bibfnamefont {C~J~C}\ \bibnamefont
  {Scott}}, \bibinfo {author} {\bibfnamefont {O~J}\ \bibnamefont {Backhouse}},
  \ and\ \bibinfo {author} {\bibfnamefont {G~H}\ \bibnamefont {Booth}},\
  }\bibfield  {title} {\enquote {\bibinfo {title} {A moment-conserving
  reformulation of {$GW$} theory},}\ }\href {https://doi.org/10.1063/5.0143291}
  {\bibfield  {journal} {\bibinfo  {journal} {The Journal of Chemical Physics}\
  }\textbf {\bibinfo {volume} {158}},\ \bibinfo {pages} {124102} (\bibinfo
  {year} {2023})}\BibitemShut {NoStop}%
\bibitem [{\citenamefont {Mertins}\ and\ \citenamefont
  {Schirmer}(1996)}]{mertins1996algebraicI}%
  \BibitemOpen
  \bibfield  {author} {\bibinfo {author} {\bibfnamefont {F}~\bibnamefont
  {Mertins}}\ and\ \bibinfo {author} {\bibfnamefont {J}~\bibnamefont
  {Schirmer}},\ }\bibfield  {title} {\enquote {\bibinfo {title} {Algebraic
  propagator approaches and intermediate-state representations. {I}. the
  biorthogonal and unitary coupled-cluster methods},}\ }\href
  {https://doi.org/10.1103/PhysRevA.53.2140} {\bibfield  {journal} {\bibinfo
  {journal} {Physical Review A}\ }\textbf {\bibinfo {volume} {53}},\ \bibinfo
  {pages} {2140} (\bibinfo {year} {1996})}\BibitemShut {NoStop}%
\bibitem [{\citenamefont {Mertins}\ \emph {et~al.}(1996)\citenamefont
  {Mertins}, \citenamefont {Schirmer},\ and\ \citenamefont
  {Tarantelli}}]{mertins1996algebraicII}%
  \BibitemOpen
  \bibfield  {author} {\bibinfo {author} {\bibfnamefont {F}~\bibnamefont
  {Mertins}}, \bibinfo {author} {\bibfnamefont {J}~\bibnamefont {Schirmer}}, \
  and\ \bibinfo {author} {\bibfnamefont {A}~\bibnamefont {Tarantelli}},\
  }\bibfield  {title} {\enquote {\bibinfo {title} {Algebraic propagator
  approaches and intermediate-state representations. {II}. the
  equation-of-motion methods for {N, N$\pm$1, and N$\pm$2 electrons}},}\ }\href
  {https://doi.org/10.1103/PhysRevA.53.2153} {\bibfield  {journal} {\bibinfo
  {journal} {Physical Review A}\ }\textbf {\bibinfo {volume} {53}},\ \bibinfo
  {pages} {2153} (\bibinfo {year} {1996})}\BibitemShut {NoStop}%
\bibitem [{\citenamefont {von Barth}\ and\ \citenamefont
  {Holm}(1996)}]{von1996self}%
  \BibitemOpen
  \bibfield  {author} {\bibinfo {author} {\bibfnamefont {U}~\bibnamefont {von
  Barth}}\ and\ \bibinfo {author} {\bibfnamefont {B}~\bibnamefont {Holm}},\
  }\bibfield  {title} {\enquote {\bibinfo {title} {Self-consistent {$GW_0$}
  results for the electron gas: Fixed screened potential {$W_0$} within the
  {Random-Phase} approximation},}\ }\href
  {https://doi.org/10.1103/PhysRevB.54.8411} {\bibfield  {journal} {\bibinfo
  {journal} {Physical Review B}\ }\textbf {\bibinfo {volume} {54}},\ \bibinfo
  {pages} {8411} (\bibinfo {year} {1996})}\BibitemShut {NoStop}%
\bibitem [{\citenamefont {Di~Sabatino}\ \emph {et~al.}(2021)\citenamefont
  {Di~Sabatino}, \citenamefont {Loos},\ and\ \citenamefont
  {Romaniello}}]{di2021scrutinizing}%
  \BibitemOpen
  \bibfield  {author} {\bibinfo {author} {\bibfnamefont {S}~\bibnamefont
  {Di~Sabatino}}, \bibinfo {author} {\bibfnamefont {P-F}\ \bibnamefont {Loos}},
  \ and\ \bibinfo {author} {\bibfnamefont {P}~\bibnamefont {Romaniello}},\
  }\bibfield  {title} {\enquote {\bibinfo {title} {Scrutinizing {$GW$}-based
  methods using the {H}ubbard dimer},}\ }\href
  {https://doi.org/10.3389/fchem.2021.751054} {\bibfield  {journal} {\bibinfo
  {journal} {Frontiers in Chemistry}\ }\textbf {\bibinfo {volume} {9}},\
  \bibinfo {pages} {751054} (\bibinfo {year} {2021})}\BibitemShut {NoStop}%
\bibitem [{\citenamefont {Ortiz}(2020)}]{ortiz2020dyson}%
  \BibitemOpen
  \bibfield  {author} {\bibinfo {author} {\bibfnamefont {JV}~\bibnamefont
  {Ortiz}},\ }\bibfield  {title} {\enquote {\bibinfo {title} {Dyson-orbital
  concepts for description of electrons in molecules},}\ }\href
  {https://doi.org/10.1063/5.0016472} {\bibfield  {journal} {\bibinfo
  {journal} {The Journal of Chemical Physics}\ }\textbf {\bibinfo {volume}
  {153}},\ \bibinfo {pages} {070902} (\bibinfo {year} {2020})}\BibitemShut
  {NoStop}%
\bibitem [{\citenamefont {Hirata}\ \emph {et~al.}(2024)\citenamefont {Hirata},
  \citenamefont {Grabowski}, \citenamefont {Ortiz},\ and\ \citenamefont
  {Bartlett}}]{hirata2024nonconvergence}%
  \BibitemOpen
  \bibfield  {author} {\bibinfo {author} {\bibfnamefont {So}~\bibnamefont
  {Hirata}}, \bibinfo {author} {\bibfnamefont {Ireneusz}\ \bibnamefont
  {Grabowski}}, \bibinfo {author} {\bibfnamefont {J~Vincent}\ \bibnamefont
  {Ortiz}}, \ and\ \bibinfo {author} {\bibfnamefont {Rodney~J}\ \bibnamefont
  {Bartlett}},\ }\bibfield  {title} {\enquote {\bibinfo {title} {Nonconvergence
  of the {Feynman-Dyson} diagrammatic perturbation expansion of propagators},}\
  }\href {https://doi.org/10.1103/PhysRevA.109.052220} {\bibfield  {journal}
  {\bibinfo  {journal} {Physical Review A}\ }\textbf {\bibinfo {volume}
  {109}},\ \bibinfo {pages} {052220} (\bibinfo {year} {2024})}\BibitemShut
  {NoStop}%
\bibitem [{\citenamefont {Scuseria}(1995)}]{scuseria1995connections}%
  \BibitemOpen
  \bibfield  {author} {\bibinfo {author} {\bibfnamefont {G~E}\ \bibnamefont
  {Scuseria}},\ }\bibfield  {title} {\enquote {\bibinfo {title} {On the
  connections between {Brueckner}--coupled-cluster, density-dependent
  {Hartree--Fock}, and {Density Functional Theory}},}\ }\href
  {https://doi.org/10.1002/qua.560550211} {\bibfield  {journal} {\bibinfo
  {journal} {International Journal of Quantum Chemistry}\ }\textbf {\bibinfo
  {volume} {55}},\ \bibinfo {pages} {165--171} (\bibinfo {year}
  {1995})}\BibitemShut {NoStop}%
\bibitem [{\citenamefont {Gauss}\ and\ \citenamefont
  {Stanton}(1995)}]{gauss1995coupled}%
  \BibitemOpen
  \bibfield  {author} {\bibinfo {author} {\bibfnamefont {J}~\bibnamefont
  {Gauss}}\ and\ \bibinfo {author} {\bibfnamefont {J~F}\ \bibnamefont
  {Stanton}},\ }\bibfield  {title} {\enquote {\bibinfo {title} {Coupled-cluster
  calculations of nuclear magnetic resonance chemical shifts},}\ }\href
  {https://doi.org/10.1063/1.470240} {\bibfield  {journal} {\bibinfo  {journal}
  {The Journal of Chemical Physics}\ }\textbf {\bibinfo {volume} {103}},\
  \bibinfo {pages} {3561--3577} (\bibinfo {year} {1995})}\BibitemShut {NoStop}%
\bibitem [{\citenamefont {Nooijen}\ and\ \citenamefont
  {Bartlett}(1995)}]{nooijen1995equation}%
  \BibitemOpen
  \bibfield  {author} {\bibinfo {author} {\bibfnamefont {M}~\bibnamefont
  {Nooijen}}\ and\ \bibinfo {author} {\bibfnamefont {R~J}\ \bibnamefont
  {Bartlett}},\ }\bibfield  {title} {\enquote {\bibinfo {title}
  {Equation-of-motion coupled-cluster method for electron attachment},}\ }\href
  {https://doi.org/10.1063/1.468592} {\bibfield  {journal} {\bibinfo  {journal}
  {The Journal of Chemical Physics}\ }\textbf {\bibinfo {volume} {102}},\
  \bibinfo {pages} {3629--3647} (\bibinfo {year} {1995})}\BibitemShut {NoStop}%
\bibitem [{\citenamefont {Nooijen}\ and\ \citenamefont
  {Bartlett}(1997)}]{nooijen1997similarity}%
  \BibitemOpen
  \bibfield  {author} {\bibinfo {author} {\bibfnamefont {M}~\bibnamefont
  {Nooijen}}\ and\ \bibinfo {author} {\bibfnamefont {R~J}\ \bibnamefont
  {Bartlett}},\ }\bibfield  {title} {\enquote {\bibinfo {title} {Similarity
  transformed equation-of-motion coupled-cluster theory: Details, examples, and
  comparisons},}\ }\href {https://doi.org/10.1063/1.474922} {\bibfield
  {journal} {\bibinfo  {journal} {The Journal of Chemical Physics}\ }\textbf
  {\bibinfo {volume} {107}},\ \bibinfo {pages} {6812--6830} (\bibinfo {year}
  {1997})}\BibitemShut {NoStop}%
\bibitem [{\citenamefont {Musia{\l}}\ \emph {et~al.}(2003)\citenamefont
  {Musia{\l}}, \citenamefont {Kucharski},\ and\ \citenamefont
  {Bartlett}}]{musial2003equation}%
  \BibitemOpen
  \bibfield  {author} {\bibinfo {author} {\bibfnamefont {M}~\bibnamefont
  {Musia{\l}}}, \bibinfo {author} {\bibfnamefont {S~A}\ \bibnamefont
  {Kucharski}}, \ and\ \bibinfo {author} {\bibfnamefont {R~J}\ \bibnamefont
  {Bartlett}},\ }\bibfield  {title} {\enquote {\bibinfo {title}
  {Equation-of-motion coupled-cluster method with full inclusion of the
  connected triple excitations for ionized states: {IP-EOM-CCSDT}},}\ }\href
  {https://doi.org/10.1063/1.1527013} {\bibfield  {journal} {\bibinfo
  {journal} {The Journal of Chemical Physics}\ }\textbf {\bibinfo {volume}
  {118}},\ \bibinfo {pages} {1128--1136} (\bibinfo {year} {2003})}\BibitemShut
  {NoStop}%
\bibitem [{\citenamefont {Shavitt}\ and\ \citenamefont
  {Bartlett}(2009)}]{shavitt2009many}%
  \BibitemOpen
  \bibfield  {author} {\bibinfo {author} {\bibfnamefont {I}~\bibnamefont
  {Shavitt}}\ and\ \bibinfo {author} {\bibfnamefont {R~J}\ \bibnamefont
  {Bartlett}},\ }\href@noop {} {\emph {\bibinfo {title} {Many-body methods in
  chemistry and physics: MBPT and coupled-cluster theory}}}\ (\bibinfo
  {publisher} {Cambridge university press},\ \bibinfo {year}
  {2009})\BibitemShut {NoStop}%
\bibitem [{\citenamefont {Scuseria}\ \emph {et~al.}(2008)\citenamefont
  {Scuseria}, \citenamefont {Henderson},\ and\ \citenamefont
  {Sorensen}}]{scuseria2008ground}%
  \BibitemOpen
  \bibfield  {author} {\bibinfo {author} {\bibfnamefont {G~E}\ \bibnamefont
  {Scuseria}}, \bibinfo {author} {\bibfnamefont {T~M}\ \bibnamefont
  {Henderson}}, \ and\ \bibinfo {author} {\bibfnamefont {D~C}\ \bibnamefont
  {Sorensen}},\ }\bibfield  {title} {\enquote {\bibinfo {title} {The ground
  state correlation energy of the {Random-Phase} approximation from a ring
  coupled-cluster doubles approach},}\ }\href
  {https://doi.org/10.1063/1.3043729} {\bibfield  {journal} {\bibinfo
  {journal} {The Journal of Chemical Physics}\ }\textbf {\bibinfo {volume}
  {129}},\ \bibinfo {pages} {231101} (\bibinfo {year} {2008})}\BibitemShut
  {NoStop}%
\bibitem [{\citenamefont {Scuseria}\ \emph {et~al.}(2013)\citenamefont
  {Scuseria}, \citenamefont {Henderson},\ and\ \citenamefont
  {Bulik}}]{scuseria2013particle}%
  \BibitemOpen
  \bibfield  {author} {\bibinfo {author} {\bibfnamefont {G~E}\ \bibnamefont
  {Scuseria}}, \bibinfo {author} {\bibfnamefont {T~M}\ \bibnamefont
  {Henderson}}, \ and\ \bibinfo {author} {\bibfnamefont {I~W}\ \bibnamefont
  {Bulik}},\ }\bibfield  {title} {\enquote {\bibinfo {title} {Particle-particle
  and quasiparticle {Random-Phase} approximations: Connections to
  coupled-cluster theory},}\ }\href {https://doi.org/10.1063/1.4820557}
  {\bibfield  {journal} {\bibinfo  {journal} {The Journal of Chemical Physics}\
  }\textbf {\bibinfo {volume} {139}},\ \bibinfo {pages} {104113} (\bibinfo
  {year} {2013})}\BibitemShut {NoStop}%
\bibitem [{\citenamefont {Rishi}\ \emph {et~al.}(2020)\citenamefont {Rishi},
  \citenamefont {Perera},\ and\ \citenamefont {Bartlett}}]{rishi2020route}%
  \BibitemOpen
  \bibfield  {author} {\bibinfo {author} {\bibfnamefont {V}~\bibnamefont
  {Rishi}}, \bibinfo {author} {\bibfnamefont {A}~\bibnamefont {Perera}}, \ and\
  \bibinfo {author} {\bibfnamefont {R~J}\ \bibnamefont {Bartlett}},\ }\bibfield
   {title} {\enquote {\bibinfo {title} {A route to improving {RPA} excitation
  energies through its connection to equation-of-motion coupled-cluster
  theory},}\ }\href {https://doi.org/10.1063/5.0023862} {\bibfield  {journal}
  {\bibinfo  {journal} {The Journal of Chemical Physics}\ }\textbf {\bibinfo
  {volume} {153}},\ \bibinfo {pages} {234101} (\bibinfo {year}
  {2020})}\BibitemShut {NoStop}%
\bibitem [{\citenamefont {Ring}\ and\ \citenamefont
  {Schuck}(2004)}]{ring2004nuclear}%
  \BibitemOpen
  \bibfield  {author} {\bibinfo {author} {\bibfnamefont {P}~\bibnamefont
  {Ring}}\ and\ \bibinfo {author} {\bibfnamefont {P}~\bibnamefont {Schuck}},\
  }\href@noop {} {\emph {\bibinfo {title} {The nuclear many-body problem}}}\
  (\bibinfo  {publisher} {Springer Science \& Business Media},\ \bibinfo {year}
  {2004})\BibitemShut {NoStop}%
\bibitem [{\citenamefont {Chiles}\ and\ \citenamefont
  {Dykstra}(1981)}]{chiles1981electron}%
  \BibitemOpen
  \bibfield  {author} {\bibinfo {author} {\bibfnamefont {R~A}\ \bibnamefont
  {Chiles}}\ and\ \bibinfo {author} {\bibfnamefont {C~E}\ \bibnamefont
  {Dykstra}},\ }\bibfield  {title} {\enquote {\bibinfo {title} {{An electron
  pair operator approach to coupled-cluster wave functions. Application to
  He$_2$, Be$_2$, and Mg$_2$ and comparison with CEPA methods}},}\ }\href
  {https://doi.org/10.1063/1.441643} {\bibfield  {journal} {\bibinfo  {journal}
  {The Journal of Chemical Physics}\ }\textbf {\bibinfo {volume} {74}},\
  \bibinfo {pages} {4544--4556} (\bibinfo {year} {1981})}\BibitemShut {NoStop}%
\bibitem [{\citenamefont {Handy}\ \emph {et~al.}(1989)\citenamefont {Handy},
  \citenamefont {Pople}, \citenamefont {Head-Gordon}, \citenamefont
  {Raghavachari},\ and\ \citenamefont {Trucks}}]{handy1989size}%
  \BibitemOpen
  \bibfield  {author} {\bibinfo {author} {\bibfnamefont {N~C}\ \bibnamefont
  {Handy}}, \bibinfo {author} {\bibfnamefont {J~A}\ \bibnamefont {Pople}},
  \bibinfo {author} {\bibfnamefont {M}~\bibnamefont {Head-Gordon}}, \bibinfo
  {author} {\bibfnamefont {K}~\bibnamefont {Raghavachari}}, \ and\ \bibinfo
  {author} {\bibfnamefont {G~W}\ \bibnamefont {Trucks}},\ }\bibfield  {title}
  {\enquote {\bibinfo {title} {Size-consistent {Brueckner} theory limited to
  double substitutions},}\ }\href
  {https://doi.org/10.1016/0009-2614(89)85013-4} {\bibfield  {journal}
  {\bibinfo  {journal} {Chemical Physics Letters}\ }\textbf {\bibinfo {volume}
  {164}},\ \bibinfo {pages} {185--192} (\bibinfo {year} {1989})}\BibitemShut
  {NoStop}%
\bibitem [{\citenamefont {Helgaker}\ \emph {et~al.}(2013)\citenamefont
  {Helgaker}, \citenamefont {Jorgensen},\ and\ \citenamefont
  {Olsen}}]{helgaker2013molecular}%
  \BibitemOpen
  \bibfield  {author} {\bibinfo {author} {\bibfnamefont {Trygve}\ \bibnamefont
  {Helgaker}}, \bibinfo {author} {\bibfnamefont {Poul}\ \bibnamefont
  {Jorgensen}}, \ and\ \bibinfo {author} {\bibfnamefont {Jeppe}\ \bibnamefont
  {Olsen}},\ }\href@noop {} {\emph {\bibinfo {title} {Molecular
  electronic-structure theory}}}\ (\bibinfo  {publisher} {John Wiley \& Sons},\
  \bibinfo {year} {2013})\BibitemShut {NoStop}%
\bibitem [{\citenamefont {Tew}(2016)}]{tew2016explicitly}%
  \BibitemOpen
  \bibfield  {author} {\bibinfo {author} {\bibfnamefont {D~P}\ \bibnamefont
  {Tew}},\ }\bibfield  {title} {\enquote {\bibinfo {title} {Explicitly
  correlated coupled-cluster theory with {Brueckner} orbitals},}\ }\href
  {https://doi.org/10.1063/1.4960655} {\bibfield  {journal} {\bibinfo
  {journal} {The Journal of Chemical Physics}\ }\textbf {\bibinfo {volume}
  {145}} (\bibinfo {year} {2016})}\BibitemShut {NoStop}%
\bibitem [{\citenamefont {Stanton}\ and\ \citenamefont
  {Gauss}(1999)}]{stanton1999simple}%
  \BibitemOpen
  \bibfield  {author} {\bibinfo {author} {\bibfnamefont {J~F}\ \bibnamefont
  {Stanton}}\ and\ \bibinfo {author} {\bibfnamefont {J}~\bibnamefont {Gauss}},\
  }\bibfield  {title} {\enquote {\bibinfo {title} {A simple scheme for the
  direct calculation of ionization potentials with coupled-cluster theory that
  exploits established excitation energy methods},}\ }\href
  {https://doi.org/10.1063/1.479673} {\bibfield  {journal} {\bibinfo  {journal}
  {The Journal of Chemical Physics}\ }\textbf {\bibinfo {volume} {111}},\
  \bibinfo {pages} {8785--8788} (\bibinfo {year} {1999})}\BibitemShut {NoStop}%
\bibitem [{\citenamefont {Hirata}\ \emph
  {et~al.}(2000{\natexlab{a}})\citenamefont {Hirata}, \citenamefont {Nooijen},\
  and\ \citenamefont {Bartlett}}]{hirata2000high}%
  \BibitemOpen
  \bibfield  {author} {\bibinfo {author} {\bibfnamefont {S}~\bibnamefont
  {Hirata}}, \bibinfo {author} {\bibfnamefont {M}~\bibnamefont {Nooijen}}, \
  and\ \bibinfo {author} {\bibfnamefont {R~J}\ \bibnamefont {Bartlett}},\
  }\bibfield  {title} {\enquote {\bibinfo {title} {High-order determinantal
  equation-of-motion coupled-cluster calculations for electronic excited
  states},}\ }\href {https://doi.org/10.1016/S0009-2614(00)00772-7} {\bibfield
  {journal} {\bibinfo  {journal} {Chemical Physics Letters}\ }\textbf {\bibinfo
  {volume} {326}},\ \bibinfo {pages} {255--262} (\bibinfo {year}
  {2000}{\natexlab{a}})}\BibitemShut {NoStop}%
\bibitem [{\citenamefont {Hirata}\ \emph
  {et~al.}(2000{\natexlab{b}})\citenamefont {Hirata}, \citenamefont {Nooijen},\
  and\ \citenamefont {Bartlett}}]{hirata2000high1}%
  \BibitemOpen
  \bibfield  {author} {\bibinfo {author} {\bibfnamefont {S}~\bibnamefont
  {Hirata}}, \bibinfo {author} {\bibfnamefont {M}~\bibnamefont {Nooijen}}, \
  and\ \bibinfo {author} {\bibfnamefont {R~J}\ \bibnamefont {Bartlett}},\
  }\bibfield  {title} {\enquote {\bibinfo {title} {High-order determinantal
  equation-of-motion coupled-cluster calculations for ionized and
  electron-attached states},}\ }\href
  {https://doi.org/10.1016/S0009-2614(00)00965-9} {\bibfield  {journal}
  {\bibinfo  {journal} {Chemical Physics Letters}\ }\textbf {\bibinfo {volume}
  {328}},\ \bibinfo {pages} {459--468} (\bibinfo {year}
  {2000}{\natexlab{b}})}\BibitemShut {NoStop}%
\bibitem [{\citenamefont {Hirata}(2004)}]{hirata2004higher}%
  \BibitemOpen
  \bibfield  {author} {\bibinfo {author} {\bibfnamefont {S}~\bibnamefont
  {Hirata}},\ }\bibfield  {title} {\enquote {\bibinfo {title} {Higher-order
  equation-of-motion coupled-cluster methods},}\ }\href
  {https://doi.org/10.1063/1.1753556} {\bibfield  {journal} {\bibinfo
  {journal} {The Journal of Chemical Physics}\ }\textbf {\bibinfo {volume}
  {121}},\ \bibinfo {pages} {51--59} (\bibinfo {year} {2004})}\BibitemShut
  {NoStop}%
\bibitem [{\citenamefont {Hubbard}(1964)}]{hubbard1964electron}%
  \BibitemOpen
  \bibfield  {author} {\bibinfo {author} {\bibfnamefont {J}~\bibnamefont
  {Hubbard}},\ }\bibfield  {title} {\enquote {\bibinfo {title} {Electron
  correlations in narrow energy bands {III}: An improved solution},}\ }\href
  {https://doi.org/10.1098/rspa.1964.0190} {\bibfield  {journal} {\bibinfo
  {journal} {Proceedings of the Royal Society of London. Series A. Mathematical
  and Physical Sciences}\ }\textbf {\bibinfo {volume} {281}},\ \bibinfo {pages}
  {401--419} (\bibinfo {year} {1964})}\BibitemShut {NoStop}%
\bibitem [{\citenamefont {Coveney}\ and\ \citenamefont
  {Tew}(2023)}]{coveney2023regularized}%
  \BibitemOpen
  \bibfield  {author} {\bibinfo {author} {\bibfnamefont {C~J~N}\ \bibnamefont
  {Coveney}}\ and\ \bibinfo {author} {\bibfnamefont {D~P}\ \bibnamefont
  {Tew}},\ }\bibfield  {title} {\enquote {\bibinfo {title} {A regularized
  second-order correlation method from {Green’s} function theory},}\ }\href
  {https://doi.org/10.1021/acs.jctc.3c00246} {\bibfield  {journal} {\bibinfo
  {journal} {Journal of Chemical Theory and Computation}\ }\textbf {\bibinfo
  {volume} {19}},\ \bibinfo {pages} {3915--3928} (\bibinfo {year}
  {2023})}\BibitemShut {NoStop}%
\bibitem [{\citenamefont {Romaniello}\ \emph {et~al.}(2009)\citenamefont
  {Romaniello}, \citenamefont {Guyot},\ and\ \citenamefont
  {Reining}}]{romaniello2009self}%
  \BibitemOpen
  \bibfield  {author} {\bibinfo {author} {\bibfnamefont {P}~\bibnamefont
  {Romaniello}}, \bibinfo {author} {\bibfnamefont {S}~\bibnamefont {Guyot}}, \
  and\ \bibinfo {author} {\bibfnamefont {L}~\bibnamefont {Reining}},\
  }\bibfield  {title} {\enquote {\bibinfo {title} {The self-energy beyond
  {$GW$}: Local and nonlocal vertex corrections},}\ }\href
  {https://doi.org/10.1063/1.3249965} {\bibfield  {journal} {\bibinfo
  {journal} {The Journal of Chemical Physics}\ }\textbf {\bibinfo {volume}
  {131}},\ \bibinfo {pages} {154111} (\bibinfo {year} {2009})}\BibitemShut
  {NoStop}%
\bibitem [{\citenamefont {Romaniello}\ \emph {et~al.}(2012)\citenamefont
  {Romaniello}, \citenamefont {Bechstedt},\ and\ \citenamefont
  {Reining}}]{romaniello2012beyond}%
  \BibitemOpen
  \bibfield  {author} {\bibinfo {author} {\bibfnamefont {P}~\bibnamefont
  {Romaniello}}, \bibinfo {author} {\bibfnamefont {F}~\bibnamefont
  {Bechstedt}}, \ and\ \bibinfo {author} {\bibfnamefont {L}~\bibnamefont
  {Reining}},\ }\bibfield  {title} {\enquote {\bibinfo {title} {Beyond the
  {$GW$} approximation: Combining correlation channels},}\ }\href
  {https://doi.org/10.1103/PhysRevB.85.155131} {\bibfield  {journal} {\bibinfo
  {journal} {Physical Review B}\ }\textbf {\bibinfo {volume} {85}},\ \bibinfo
  {pages} {155131} (\bibinfo {year} {2012})}\BibitemShut {NoStop}%
\bibitem [{\citenamefont {Lani}\ \emph {et~al.}(2012)\citenamefont {Lani},
  \citenamefont {Romaniello},\ and\ \citenamefont
  {Reining}}]{lani2012approximations}%
  \BibitemOpen
  \bibfield  {author} {\bibinfo {author} {\bibfnamefont {G}~\bibnamefont
  {Lani}}, \bibinfo {author} {\bibfnamefont {P}~\bibnamefont {Romaniello}}, \
  and\ \bibinfo {author} {\bibfnamefont {L}~\bibnamefont {Reining}},\
  }\bibfield  {title} {\enquote {\bibinfo {title} {Approximations for many-body
  {Green's} functions: insights from the fundamental equations},}\ }\href
  {https://iopscience.iop.org/article/10.1088/1367-2630/14/1/013056/meta}
  {\bibfield  {journal} {\bibinfo  {journal} {New Journal of Physics}\ }\textbf
  {\bibinfo {volume} {14}},\ \bibinfo {pages} {013056} (\bibinfo {year}
  {2012})}\BibitemShut {NoStop}%
\bibitem [{\citenamefont {Martin}\ \emph {et~al.}(2016)\citenamefont {Martin},
  \citenamefont {Reining},\ and\ \citenamefont
  {Ceperley}}]{martin2016interacting}%
  \BibitemOpen
  \bibfield  {author} {\bibinfo {author} {\bibfnamefont {R~M}\ \bibnamefont
  {Martin}}, \bibinfo {author} {\bibfnamefont {L}~\bibnamefont {Reining}}, \
  and\ \bibinfo {author} {\bibfnamefont {D~M}\ \bibnamefont {Ceperley}},\
  }\href@noop {} {\emph {\bibinfo {title} {Interacting Electrons}}}\ (\bibinfo
  {publisher} {Cambridge University Press},\ \bibinfo {year}
  {2016})\BibitemShut {NoStop}%
\bibitem [{\citenamefont {Riva}\ \emph {et~al.}(2023)\citenamefont {Riva},
  \citenamefont {Romaniello},\ and\ \citenamefont
  {Berger}}]{riva2023multichannel}%
  \BibitemOpen
  \bibfield  {author} {\bibinfo {author} {\bibfnamefont {G}~\bibnamefont
  {Riva}}, \bibinfo {author} {\bibfnamefont {P}~\bibnamefont {Romaniello}}, \
  and\ \bibinfo {author} {\bibfnamefont {J~A}\ \bibnamefont {Berger}},\
  }\bibfield  {title} {\enquote {\bibinfo {title} {Multichannel {Dyson}
  equation: Coupling many-body {Green’s} functions},}\ }\href
  {https://doi.org/10.1103/PhysRevLett.131.216401} {\bibfield  {journal}
  {\bibinfo  {journal} {Physical Review Letters}\ }\textbf {\bibinfo {volume}
  {131}},\ \bibinfo {pages} {216401} (\bibinfo {year} {2023})}\BibitemShut
  {NoStop}%
\bibitem [{\citenamefont {Riva}\ \emph {et~al.}(2024)\citenamefont {Riva},
  \citenamefont {Romaniello},\ and\ \citenamefont
  {Berger}}]{riva2024derivation}%
  \BibitemOpen
  \bibfield  {author} {\bibinfo {author} {\bibfnamefont {G}~\bibnamefont
  {Riva}}, \bibinfo {author} {\bibfnamefont {P}~\bibnamefont {Romaniello}}, \
  and\ \bibinfo {author} {\bibfnamefont {J~A}\ \bibnamefont {Berger}},\
  }\bibfield  {title} {\enquote {\bibinfo {title} {Derivation and analysis of
  the multichannel {Dyson} equation},}\ }\href
  {https://journals.aps.org/prb/abstract/10.1103/PhysRevB.110.115140}
  {\bibfield  {journal} {\bibinfo  {journal} {Physical Review B}\ }\textbf
  {\bibinfo {volume} {110}},\ \bibinfo {pages} {115140} (\bibinfo {year}
  {2024})}\BibitemShut {NoStop}%
\bibitem [{\citenamefont {Riva}\ \emph {et~al.}(2025)\citenamefont {Riva},
  \citenamefont {Fischer}, \citenamefont {Paggi}, \citenamefont {Berger},\ and\
  \citenamefont {Romaniello}}]{riva2025multichannel}%
  \BibitemOpen
  \bibfield  {author} {\bibinfo {author} {\bibfnamefont {G}~\bibnamefont
  {Riva}}, \bibinfo {author} {\bibfnamefont {T}~\bibnamefont {Fischer}},
  \bibinfo {author} {\bibfnamefont {S}~\bibnamefont {Paggi}}, \bibinfo {author}
  {\bibfnamefont {J~A}\ \bibnamefont {Berger}}, \ and\ \bibinfo {author}
  {\bibfnamefont {P}~\bibnamefont {Romaniello}},\ }\bibfield  {title} {\enquote
  {\bibinfo {title} {{Multichannel Dyson equations for even-and odd-order
  Green's functions: Application to double excitations}},}\ }\href
  {https://doi.org/10.1103/PhysRevB.111.195133} {\bibfield  {journal} {\bibinfo
   {journal} {Physical Review B}\ }\textbf {\bibinfo {volume} {111}},\ \bibinfo
  {pages} {195133} (\bibinfo {year} {2025})}\BibitemShut {NoStop}%
\bibitem [{\citenamefont {Ammar}\ \emph {et~al.}(2024)\citenamefont {Ammar},
  \citenamefont {Marie}, \citenamefont {Rodr{\'\i}guez-Mayorga}, \citenamefont
  {Burton},\ and\ \citenamefont {Loos}}]{ammar2024can}%
  \BibitemOpen
  \bibfield  {author} {\bibinfo {author} {\bibfnamefont {A}~\bibnamefont
  {Ammar}}, \bibinfo {author} {\bibfnamefont {A}~\bibnamefont {Marie}},
  \bibinfo {author} {\bibfnamefont {M}~\bibnamefont {Rodr{\'\i}guez-Mayorga}},
  \bibinfo {author} {\bibfnamefont {H~G~A}\ \bibnamefont {Burton}}, \ and\
  \bibinfo {author} {\bibfnamefont {P-F}\ \bibnamefont {Loos}},\ }\bibfield
  {title} {\enquote {\bibinfo {title} {Can {$GW$} handle multireference
  systems?}}\ }\href {https://doi.org/10.1063/5.0196561} {\bibfield  {journal}
  {\bibinfo  {journal} {The Journal of Chemical Physics}\ }\textbf {\bibinfo
  {volume} {160}},\ \bibinfo {pages} {114101} (\bibinfo {year}
  {2024})}\BibitemShut {NoStop}%
\end{thebibliography}%

\clearpage

\begin{figure}
    \centering
    \includegraphics[width=8.25cm,height=3.00cm]{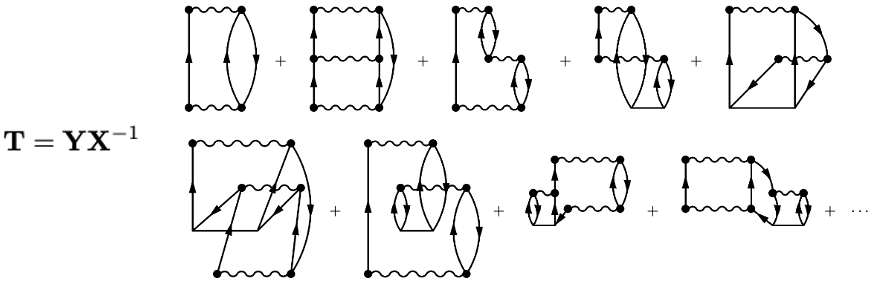}
    \caption{TOC Graphic}
    \label{fig:toc}
\end{figure}

\end{document}